%% file: SGWB_421_draft_V1.tex
\documentclass[aps,prd,twocolumn,floatfix,preprintnumbers,superscriptaddress,nofootinbib]{revtex4-1}
\usepackage{style}
\begin{document}

\title{Inflation, Proton Decay and Gravitational Waves from Metastable Strings in \texorpdfstring{$SU(4)_C \times SU(2)_L \times U(1)_R$}{421}  Model}
\input{authors.tex}
\noaffiliation

\begin{abstract}
We present a realistic supersymmetric $\mu$-hybrid inflation model within the framework of $SU(4)_C \times SU(2)_L \times U(1)_R$ gauge symmetry, wherein the symmetry breaking $SU(4)_C \times SU(2)_L \times U(1)_R\rightarrow SU(3)_C\times SU(2)_L \times U(1)_{B-L}\times U(1)_R$ occurs before observable inflation, effectively eliminating topologically stable primordial monopoles. Subsequent breaking of $U(1)_{B-L} \times U(1)_R \rightarrow U(1)_Y$ after inflation leads to the formation of superheavy metastable cosmic strings (CSs), capable of producing a stochastic gravitational wave background (SGWB) consistent with the recent PTA data. Moreover, the scalar spectral index $n_s$ and the tensor-to-scalar ratio $r$ align with Planck 2018 observations. A consistent scenario for reheating and non-thermal leptogenesis is employed to explain the observed matter content of the universe. Finally, the embedding of $G_{421}$ into the Pati-Salam gauge symmetry $G_{422}$ is briefly discussed, predicting potentially observable proton decay rates detectable at facilities such as Hyper Kamiokande and DUNE.

\end{abstract}

\maketitle

\newpage
\section{Introduction}
Gravitational waves (GWs) are ripples in the fabric of spacetime that can reveal fundamental aspects of the universe and offer a unique way to explore fundamental physics. The International Pulsar Timing Array (IPTA) collaboration found strong evidence of an isotropic stochastic GW background with frequencies in the nanohertz range \cite{NANOGrav:2023gor, EPTA:2023fyk, Reardon:2023gzh, Xu:2023wog}. While this stochastic GW background could originate from the combined effects of GWs generated by the merging of supermassive black holes in the universe \cite{NANOGrav:2023hfp}, there is also the possibility of new physics explaining such a signal \cite{NANOGrav:2023hvm}.

Grand unified theories, based on models such as $SU(4)_C \times SU(2)_L \times SU(2)_R$ \cite{Pati:1974yy} and $SO(10)$ \cite{Georgi:1974my}, present compelling extensions beyond the Standard Model of particle physics. Among the attractive features are the unification of gauge couplings and Yukawa interactions, achieved by organizing Standard Model matter into larger multiplets.
Furthermore, these models require the existence of right-handed neutrinos \cite{Gell-Mann:1979vob}, which subsequently explain the tiny masses of neutrinos \cite{Fukugita:1986hr}, consistent with observations of neutrino oscillations \cite{Magg:1980ut}. Notably, the presence of superheavy 't Hooft-Polyakov monopoles \cite{tHooft:1974kcl, Polyakov:1974ek} emerges as a generic feature within these models. Depending on the gauge symmetry-breaking patterns leading to the Standard Model, the prediction of both topologically stable \cite{Kibble:1976sj} and meta-stable cosmic strings \cite{Turok:1989ai} arises.

The recent data from Pulsar Timing Arrays (PTAs) \cite{NANOGrav:2023gor, EPTA:2023fyk, Reardon:2023gzh, Xu:2023wog} and the LIGO O3 run have imposed limitations on cosmic string properties. Specifically, to align with observations, the dimensionless string tension parameter for stable cosmic strings, denoted as $G\mu$, must be below $10^{-10}$ for PTAs and less than $10^{-7}$ for LIGO O3. For metastable strings, which contribute to a stochastic gravitational background consistent with recent PTA data, the constraint on $G\mu$ lies within the range $4.33 \times 10^{-8} \lesssim G\mu \lesssim 1.33 \times 10^{-7}$, accompanied by a metastability factor $\sqrt{\kappa_{m}} \simeq 8$ \cite{NANOGrav:2023hvm}.

In this paper, we explore how the desired metastable cosmic strings manifest in the $\mu$-hybrid inflation model \cite{Okada:2015vka} based on realistic GUT gauge symmetry, $G_{421} \equiv SU(4)_C \times SU(2)_L \times U(1)_R$. This model possesses several attractive features. The MSSM $\mu$ problem is elegantly solved by including a trilinear coupling $S h_u h_d$, which yields the MSSM $\mu$-term after the scalar component of $S$ acquires a VEV proportional to the gravitino mass $m_{3/2}$ \cite{Dvali:1997uq}. The $U(1)_R$ symmetry not only forbids several dangerous proton decay operators, but its unbroken $Z_2$ subgroup acts as a matter parity, implying a stable LSP and thus a potential dark matter candidate. Furthermore, the model yields naturally tiny neutrino masses through the seesaw mechanism and incorporates a realistic scenario of reheating and non-thermal leptogenesis to generate the observed baryon asymmetry of the universe (BAO), without leading to the gravitino overproduction problem. The breaking of $SU(4)_C$ to $SU(3)_C \times U(1)_{B-L}$ at a slightly higher scale, $M_{\text{GUT}}$, produces 'red' monopoles \cite{Lazarides:2023iim}, which are inflated away during inflation. The subsequent breaking of $U(1)_{B-L} \times U(1)_R$ to $U(1)_Y$, at a scale, $M \sim M_{\text{GUT}}/2$, occurs at the end of inflation. This produces the desired metastable strings that generate a stochastic gravitational wave background. The strings eventually disappear due to the quantum tunneling of the red monopole-antimonopole pairs. 

Additionally, we discuss the embedding of $G_{421}$ into a larger Pati-Sallam group $G_{422}\equiv SU(4)_C \times SU(2)_L \times SU(2)_R$  . With the addition of minimal superfields required for this embedding, $G_{421}$ can also predict observable proton decay discussed in \cite{Lazarides:2020bgy} for the $G_{422}$  model. It is important to note that the chirality non-flipping LLRR type proton decay operators are found to play a dominant role in these predictions. These type of operators are studied in other GUTs like flipped $SU(5)$, \cite{Mehmood:2020irm, Abid:2021jvn}, $SU(5)$ model with missing doublet mechanism and GUT scale Higgs in $75$ representation \cite{Mehmood:2023gmm} and R-symmetric $SU(5)$ model with missing doublet mechanism and GUT scale Higgs in $24$ representation \cite{Ijaz:2023cvc}. In this paper, we explored the parameter range that predicts observable proton decay for next-generation experiments like Hyper Kamiokande \cite{Hyper-Kamiokande:2018ofw} and DUNE \cite{DUNE:2020ypp,DUNE:2015lol}.

The layout of this paper is as follows: In \cref{model}, we provide an overview of the key characteristics of the $ SU(4)_C\times SU(2)_L\times U(1)_R$ model, including symmetry breaking and the evolution of gauge couplings. \Cref{model1} explores $\mu$-hybrid inflation scenario, reheating and leptogenesis. The results of the numerical analysis are displayed in   \cref{Nans}, including a discussion of the stochastic gravitational waves parameter space consistent with NANOGrav and other experiments. In \cref{sec2}, we discuss the possible embedding of $421$ in the Pati-Salam model and proton decay. Our conclusions are summarized in \cref{sec8}.


\section{Supersymmetric \texorpdfstring{$SU(4)_c \times SU(2)_L \times U(1)_R$}{422} Model}\label{model}
\begin{table*}[t!]
\caption{The decomposition of matter and Higgs representations of $G_{421}\equiv SU(4)_C \times SU(2)_L \times U(1)_R$ under $G_{3211}\equiv SU(3)_C \times SU(2)_L  \times U(1)_R\times U(1)_{B-L}$ and SM $G_{321}\equiv SU(3)_C \times SU(2)_L   \times U(1)_Y$ along with their global $U(1)_{R'}$ charges.\label{content}}
    \begin{ruledtabular}
        \begin{tabular}{cccllc}
            &$G_{421}$&
            $q\bigl(U(1)_{R'}\bigr)$&
            \qquad$  G_{3211}$&
            \qquad$  G_{321}$&\\
            \hline
            &$F_i \left(4,2, 0 \right)$  &  1/2 &       $Q_i \left(3, 2, 0, 1/3 \right)$ &         $Q_i \left( 3,2, 1/6 \right)$ & \\
			  && & $L_i\left(1,2, 0, -1\right)$   &      $L_i\left(1,2,-1/2\right)$ &  \\
			&{$F^{c}_{ui} \left( \bar{4},1, -1/2 \right)$}     &  $1/2$   &     $u^c_i\left( \bar{3}, 1, -1/2, -1/3 \right)$ & $u^c_i\left( \bar{3}, 1, -2/3 \right)$ &    \\
			&&    &     $\nu^c_i\left(1, 1, -1/2, 1 \right)$ & $\nu^c_i\left( 1,1, 0 \right)$    & \\
			&{$F^{c}_{di} \left( \bar{4}, 1, 1/2 \right)$} &  $1/2$    &         $d^c_i\left( \bar{3}, 1, 1/2, -1/3 \right)$  & $d^c_i\left( \bar{3}, 1, 1/3 \right)$ &\\
			&& & $e^c_i\left(1, 1, 1/2, 1 \right)$ & $e^c_i\left(1, 1, 1 \right)$& \\
            \hline 
			&{$\Phi \left(15,1 , 0\right)$}	&   $1/5$    & $\Phi_{0}(1,1, 0, 0)$ & $\Phi_{0}(1,1, 0)$ &\\
			  & & &     $G_{\Phi} (8,1, 0, 0)$      &     $G_{\Phi} (8,1, 0)$  & \\
			  & & &  $X_{\Phi} (3,1, 0, 2/3)$      &     $X_{\Phi} (3,1, 1/3)$ &  \\
			  & &  & $Y_{\Phi} (\bar{3},1, 0, -2/3)$      &      $Y_{\Phi} (\bar{3},1, -1/3)$  & \\
			&{$H \left(4, 1, 1/2\right)$}	& $0$     & $u_H(3, 1, 1/2, 1/3)$   & $u_H(3, 1, 2/3)$& \\
			  & & &  $\nu_H (1,1, 1/2, -1)$    &      $\nu_H (1, 1, 0)$ &  \\
			  &{$\overline{H} \left(\bar{4},1,-1/2\right)$}	&     $0$     & $\bar{u}_H(\bar{3}, 1, -1/2, -1/3)$   & $\bar{u}_H(\bar{3}, 1, -2/3)$& \\
			   && &  $\bar{\nu}_{H} (1, 1, -1/2, 1)$    &      $\bar{\nu}_{H} (1,1, 0)$ &  \\
		    &{$h_u \left(1, 2, 1/2\right)$}	&   $0$     & $h_u \left( 1, 1, 1/2, 0\right)$  & $h_u \left( 1,1, 1/2\right)$&\\
			  &{$h_d \left(1, 2, -1/2\right)$} & $0$   &   $h_d \left( 1, 1, -1/2, 0\right)$  & $h_u \left( 1, 1, -1/2\right)$ &\\
			  &{$S \left( 1, 1, 0\right)$}	&     $1$        & $S \left(1, 1, 0, 0\right)$   & $S \left( 1, 1, 0\right)$&\\
        \end{tabular}
    \end{ruledtabular}
\end{table*}
The superfields containing standard model (SM) matter and Higgs content and their $U(1)_{R'}$ charges are given in \cref{content}. The $G_{421} \times U(1)_{R'}$ symmetric  superpotential is written as,
\begin{align}\label{superpot-shift}
 W& = \kappa S \left( M^2 - \, H\, \overline{H} \right)   
 +\lambda_h\, S \, h_u\, h_d
  \nonumber \\
 &+ \lambda^{(u,\nu)}_{ij} F_i F^c_{u\ j} h_u
 +\lambda^{(d,e)}_{ij} F_i F^c_{d\ j}h_d   \nonumber \\
 & +\frac{\gamma_1}{m_P}(F_u^c H)^2  +
 \frac{\gamma_2}{m_P}(F_d^c  \overline{H})^2 +\cdots \nonumber \\
 &+ W_{\Phi},
\end{align}
where $S$ is a gauge singlet superfield, $m_P$ is the reduced Planck mass, and $\left(h_u,h_d\right)$ are the higgs doublet superfields. The second line in $W$ contains Yukawa terms for quarks and leptons. The global vacuum expectation value (vev) of the relevant fields are given by,
\begin{equation}
\langle S \rangle = 0, \, \langle H \overline{H} \rangle = M^2,\,  \langle h_u \rangle = 0, \, \langle h_d \rangle = 0. 
\end{equation}
However, due to soft SUSY breaking terms, the $S$ field acquires a nonzero vev, $\langle S \rangle \sim m_{3/2} / \kappa$ \cite{Dvali:1997uq}. This leads to an effective $\mu$ term, $\mu h_u h_d $, with $\mu \sim (\lambda_h/\kappa)m_{3/2}$, thus resolving the MSSM $\mu$ problem. The triplets in $H$ and $\overline{H}$ thus obtain masses of order $m_{3/2}$. As is discussed in the next section, this $\mu$ term not only contributes during inflation but is also important in the reheating process after inflation. For relevant papers on $\mu$-hybrid inflation, see \cite{Okada:2015vka,Rehman:2017gkm,Okada:2017rbf,Lazarides:2020zof,Afzal:2022vjx,Afzal:2023cyp,Zubair:2024quc}.

The last part of the above superpotential, $W_{\Phi}$, is relevant for the $SU(4)_c$ breaking, expressed as:
\begin{eqnarray}
W_{\Phi} & = & \frac{1}{m_P^2}  \left( M_{15} \Phi^4 + \lambda_{15} \Phi^5 \right),
\end{eqnarray}
where the $\Phi$ field carries an $R'$-charge of $1/5$. Note that the first term in $W_{\Phi}$ breaks the $R'$-symmetry and could arise from a large vev of a gauge singlet field, denoted as $\langle X \rangle \sim M_{15}$ \cite{Dine:1987xk, Atick:1987gy, Dine:1987gj}, in the hidden sector while carrying an $R'$-charge of $1/5$.
The $SU(4)_C$ symmetry breaks into $SU(3)_C \times U(1)_{B-L}$ at scale $v_{m}$ by acquiring a nonzero vev $v_m$ in the color singlet direction of the adjoint representation $\Phi$, 
\begin{eqnarray}
SU(4)_c\xrightarrow{\langle\Phi\rangle_{(1,1,0)}} SU(3)_C \times U(1)_{\text{B-L}},
\end{eqnarray}
with,
\begin{equation}\label{globalvev}
  \langle\Phi\rangle_{(1,1,0)}=  \dfrac{v_m}{2\sqrt{6}}   \begin{pmatrix}
   1 & 0 & 0 & 0 \\
   0 & 1 & 0 & 0 \\
   0 & 0 & 1 & 0 \\
   0 & 0 & 0 & -3
   \end{pmatrix},
\end{equation}
where $v_m \sim M_{15}/\lambda_{15}$. This breaking creates monopoles carrying $B-L$ and color magnetic charge, which are subsequently diluted during inflation. It is assumed that this breaking happens before the observable inflation as discussed later. An alternative approach, as proposed by \cite{Afzal:2023cyp}, involves implementing a shifted $\mu$-hybrid inflation to circumvent the monopole issue.

To further break $U(1)_{B-L} \times U(1)_R$ into $U(1)_Y$ at $v_{\text{str}}$ we consider $H_{(1,1,0)}$ and $\overline{H}_{(1,1,0)}$,
\begin{eqnarray}
U(1)_{B-L} \times U(1)_R \xrightarrow{\langle H\rangle_{(1,1,0)}=\langle\overline{H}\rangle_{(1,1,0)}} U(1)_Y 
\end{eqnarray}
The SM hypercharge is given by, 
\begin{eqnarray}
Y=\sqrt{\frac{3}{5}}Y_R+\sqrt{\frac{2}{5}}(B-L),
\end{eqnarray}
where $Y_R$ is the hypercharge associated with $U(1)_R$ and We have chosen the normalization of $Y_R$ assuming $U(1)_R \supset SU(2)_R$. 

\begin{figure}[t!]
    \centering
    \includegraphics[width=1.0\linewidth]{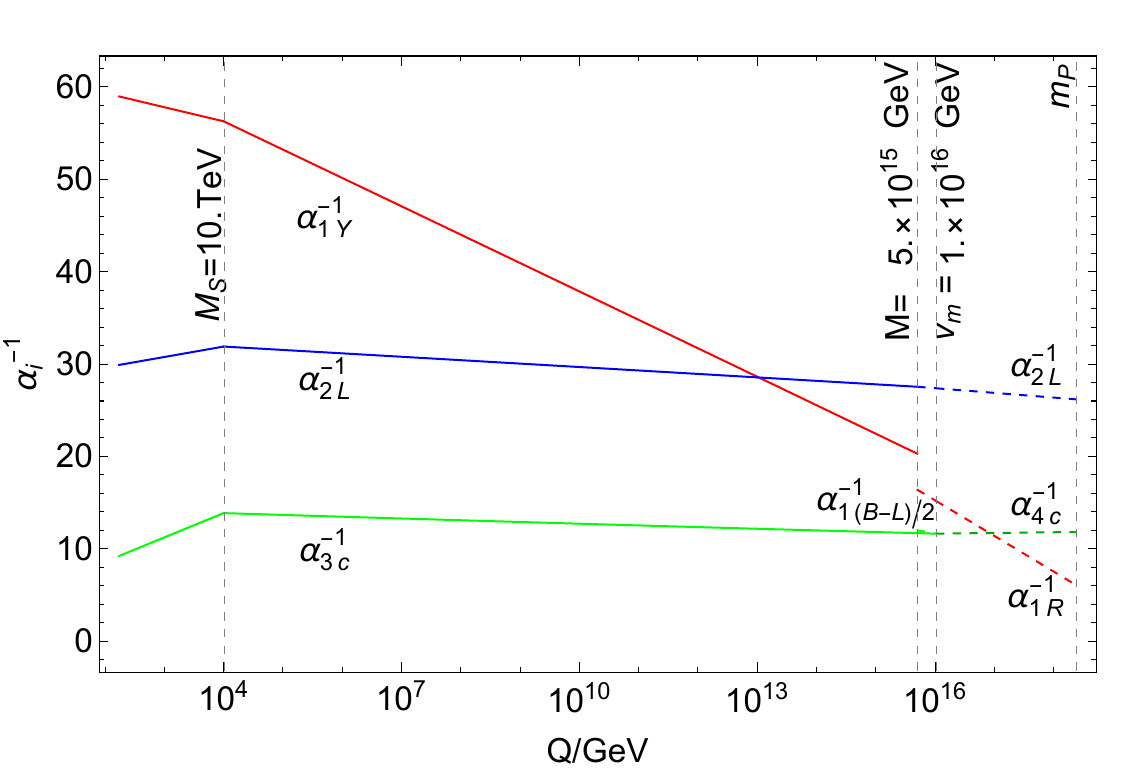}
    \caption{\label{gce}Evolution of two loop RGEs of gauge couplings for gauge groups $G_{421}$, $G_{3211}$ and $G_{321}$.}
\end{figure}

The two loop gauge coupling evolution for the gauge symmetries $SU(4)_C \times SU(2)_L \times U(1)_R$, $SU(3)_C\times SU(2)_L \times U(1)_R \times U(1)_{B-L}$ and $SU(3)_C\times SU(2)_L \times U(1)_Y$ are shown in \cref{gce}. 

\section{\texorpdfstring{$\mu$}{mu}-Hybrid Inflation}\label{model1}

The superpotential term relevant for $\mu$-hybrid inflation is \cite{Copeland:1994vg,Dvali:1994ms},
\begin{equation}
W = \kappa S(M^{2} - H \overline{H})  + \lambda_h\, S \, h_u\, h_d,
\label{superpot}
\end{equation} 
with the corresponding global susy minimum described in Eq.~(2).
The global SUSY F-term scalar potential is given by,
\begin{eqnarray}\label{VF1}
V_{F} &=& \kappa^2\,\vert M^2 - \phi \,\overline{\phi} + \, \gamma_h h_u\, h_d\vert^2 + \kappa^2 \vert s \vert^2 (\vert \phi \vert^2+\vert \overline{\phi} \vert^2)\nonumber \\
&+& \lambda_h^2 \vert s \vert^2 (\vert h_u \vert^2 +\vert h_d \vert^2) ,    
\end{eqnarray}
where $s, \,\phi , \overline{\phi }, \, h_u, \, h_d$  represent the bosonic components of the superfields $S, \, H, \overline{H}, h_u, \, h_d$ respectively. The large $s$ value during inflation provides the large masses to  
fields, $H, \overline{H}, h_u, \, h_d$. Subsequently, these fields are stabilized at zero during inflation and the tree-level global SUSY potential becomes flat $V = V_0 = \kappa^2 M^4$. As described below, the various important contributions to the scalar potential provide the necessary slope for the realization of inflation in this otherwise flat trajectory. 

The F-term supergravity (SUGRA) scalar potential is given by,
\begin{equation}
V_{F}=e^{K/m_{P}^{2}}\left(
K_{ij}^{-1}D_{z_{i}}WD_{z^{*}_j}W^{*}-3m_{P}^{-2}\left| W\right| ^{2}\right),
\label{VF}
\end{equation}
where, 
\begin{equation*}
D_{z_{i}}W \equiv \frac{\partial W}{\partial z_{i}}+m_{P}^{-2}\frac{\partial K}{\partial z_{i}}W, \,\,\,\,\,\, K_{ij} \equiv \frac{\partial ^{2}K}{\partial z_{i}\partial z_{j}^{*}},
\end{equation*}
$D_{z_{i}^{*}}W^{*}=\left( D_{z_{i}}W\right) ^{*}$, $z_i \in \{s,\phi , \overline{\phi } \}$, $z_i^*$ being conjugate, and $m_{P} \simeq 2.4\times 10^{18}$~GeV is the reduced Planck mass. In the present paper, we employ the following form of the K\"ahler potential including relevant non-minimal terms,
\begin{equation}
K \supset  |S|^{2}+ |H|^{2} + |\overline{H}|^{2}+\kappa_S \frac{|S|^{4}}{4 m_P^2} + \kappa_{SS} \frac{|S|^{6}}{6m_P^4},  \label{kahler}
\end{equation}
and the SUGRA corrections can now be calculated using the above definitions. 
Along the inflationary trajectory, SUSY is broken due to the non-zero vacuum term, $V_0$. This generates a mass splitting between the fermionic and the bosonic components of the relevant superfields and leads to radiative corrections in the scalar potential \cite{Dvali:1994ms}. Another important contribution in the scalar potential arises from the soft SUSY breaking terms \cite{Senoguz:2004vu,Buchmuller:2000zm,Rehman:2009nq}.
\par 
Including the leading order SUGRA corrections, one-loop radiative corrections and the soft SUSY breaking terms, the scalar potential along the inflationary trajectory (i.e. $\phi=0=\overline{\phi}$) can be written as \cite{Rehman:2009nq,urRehman:2006hu,Rehman:2010wm},
\begin{eqnarray}\label{scalpot}
V  &\simeq &
V_0 \left[ 1 +\frac{\kappa ^{2} \mathcal{N}}{8\pi ^{2}}F(x) + \frac{\lambda_h^{2}}{4\pi ^{2}}F(\sqrt{\gamma_h}\,x) + a \left( \frac{m_{3/2}\,x}{\kappa\,M}\right)  \right. \nonumber \\
&+&\left.  \left( \frac{m_{3/2} x}{\kappa M}\right)^2- \kappa_S  \left( \frac{M}{m_P}\right)^2 x^2 + \frac{\gamma_S}{2} \left( \frac{M}{m_P}\right)^4 x^4\right],
\end{eqnarray}
where $x=|s|/M$, $\gamma_S = 1-7\kappa_S/2 +2 \kappa_S^2 -3\kappa_{SS}$, $\gamma_h = \lambda_h / \kappa$, and $m_{3/2}$ is the gravitino mass. The one-loop radiative correction function, $F(x)$, can be expressed as
\begin{align}
F(x)&=\frac{1}{4}\left[ \left( x^{4}+1\right) \ln \frac{\left( x^{4}-1\right)}{x^{4}}+2x^{2}\ln \frac{x^{2}+1}{x^{2}-1}\right. \notag\\
&\left.+2\ln \frac{\kappa ^{2}M^{2}x^{2}}{Q^{2}}-3\right],
\end{align}
involving the renormalization scale $Q$. The radiative corrections depend on the dimensionality, $\mathcal{N}=4$, of $H,\, \overline{H}$ fields. The parameter $a$ appearing in the soft SUSY breaking linear term is defined as,
\begin{equation}
a = 2\left| 2-A\right| \cos [\arg s+\arg (2-A)].
\label{a}
\end{equation}
For simplicity, we assume $\arg s = 0$ as an initial condition corresponding to the minimum of the potential. Thus, the parameter $a$ remains constant during inflation\footnote{For a study of various possible initial conditions of $\arg s$ in standard hybrid inflation and their impact on inflationary predictions see \cite{Buchmuller:2014epa}.} \cite{urRehman:2006hu}. 

\subsection{\large{\bf Reheating and Leptogenesis}} \label{reh}
After the end of inflation, the inflaton system, composed of two complex scalar fields: $s$ and $\theta=(\delta \phi + \delta \bar{\phi})/\sqrt{2}$ with mass $m_\text{inf}$, descends toward the SUSY minimum, experiences damped oscillations about it, and eventually undergoes decay, initiating the process referred to as 'reheating'. The inflaton predominantly decays into a pair of higgsinos ($\widetilde h_u$, $\widetilde h_d$) and higgses ($ h_u$, $ h_d$), each with a decay width, $\Gamma_h$, given by \cite{Lazarides:1998qx},
\begin{equation}\label{gamma}
\Gamma_h =\Gamma(\theta \rightarrow \widetilde h_u\widetilde h_d) = \Gamma(s \rightarrow h_u h_d)=\frac{\lambda_h ^2 }{8 \pi }m_{\text{inf}},
\end{equation}
where the inflaton mass $m_\text{inf}$ is given by
	\begin{equation}
		m_\text{inf} = \sqrt{2} \kappa M.
		\label{inf_mass}
	\end{equation}
The superpotential term $(\gamma_1/m_P) \, \nu_H^2 N^{c}_i N^{c}_j$ yields the right-handed neutrino mass $M_N = \gamma_1 \left( M^2/m_P\right)$ and leads to the decay of the inflaton to a pair of right-handed neutrinos ($N$) and sneutrinos ($\widetilde{N}$) with equal decay width, given by, 
\begin{align}
\Gamma_N& = \Gamma (\theta \rightarrow NN )= \Gamma(s \rightarrow \widetilde{N}\widetilde{N} ) \notag\\
&=\dfrac{m_\text{inf}}{8\pi}\left( \frac{M_N}{M} \right)^2\left(1-\dfrac{4M_{N}^{2}}{m_\text{inf}^{2}}\right)^{1/2},
\end{align}
provided that only the lightest right-handed neutrino with mass $M_{N}$ satisfies the kinematic bound, $m_\text{inf} > 2 M_{N}$. 

With $H = 3\Gamma_\text{inf}$, we define the reheat temperature $T_R$ in terms of the inflaton decay width $\Gamma_\text{inf}$,
\begin{equation}\label{tr}
T_R=\left(\dfrac{90}{\pi^{2}g_{*}}\right)^{1/4}\sqrt{\Gamma_\text{inf} \, m_{P}},
\end{equation}
where $g_{*}= 228.75$ for MSSM. Assuming a standard thermal history, the number of e-folds, $N_{0}$, can be written in terms of the reheat temperature, $T_R$, as \cite{Liddle:2003as},
\begin{align}\label{efolds}
N_0=53+\dfrac{1}{3}\ln\left[\dfrac{T_R}{10^9 \text{ GeV}}\right]+\dfrac{2}{3}\ln\left[\dfrac{\sqrt{\kappa}\,M}{10^{15}\text{ GeV}}\right].
\end{align}

The $\Gamma_N$ channel plays a crucial role in implementing successful leptogenesis, which is partially converted into the observed baryon asymmetry through the sphaleron process \cite{Kuzmin:1985mm,Fukugita:1986hr,Khlebnikov:1988sr}. Suppression of the washout factor of lepton asymmetry can be achieved if $M_N \gg T_R$. The evaluation of the observed baryon asymmetry is expressed in terms of the lepton asymmetry factor, $\varepsilon_L$ as
\begin{align}\label{bphr}
\frac{n_{B}}{n_{\gamma}}\simeq -1.84 \,\varepsilon_L  \frac{\Gamma_N}{\Gamma_\text{inf}}\frac{T_R}{m_\text{inf}} \delta_{eff},
\end{align} 
where $\delta_{eff}$ is the CP violating phase factor and $\Gamma_\text{inf} \simeq \Gamma_h$. Assuming hierarchical neutrino masses, $\varepsilon_L$ is given by \cite{Rehman:2018gnr},
\begin{eqnarray}
(-\varepsilon_L)  \simeq \frac{3}{8\pi}  \frac{\sqrt{\Delta m_{31}^{2}} M_N}{\langle H_{u}\rangle^{2}}.
\end{eqnarray} 
Here, the atmospheric neutrino mass squared difference is $\Delta m_{31}^{2}\approx 2.6 \times 10^{-3}$ eV$^{2} $  and $\langle H_{u}\rangle \simeq 174$ GeV in the large $\tan\beta$ limit. For the observed baryon-to-photon ratio, $n_\mathrm{B} /n_\gamma = (6.12 \pm 0.04) \times 10^{-10}$ \cite{ParticleDataGroup:2020ssz}, the constraint on $|\delta_{eff}|\leq 1$ along with the kinematic bound, $m_\text{inf} \geq 2M_N$, translates into a bound on the reheat temperature, 
\begin{align}\label{lept}
T_R  \gtrsim \gamma_h^2\left(2 \times 10^7\right) \text{ GeV} \gtrsim 8 \times 10^7 \text{ GeV},
\end{align} 
were we have set $\gamma_h = 2$ in our numerical calculations.

\begin{figure*}[ht!]
\centering\includegraphics[width=\textwidth]{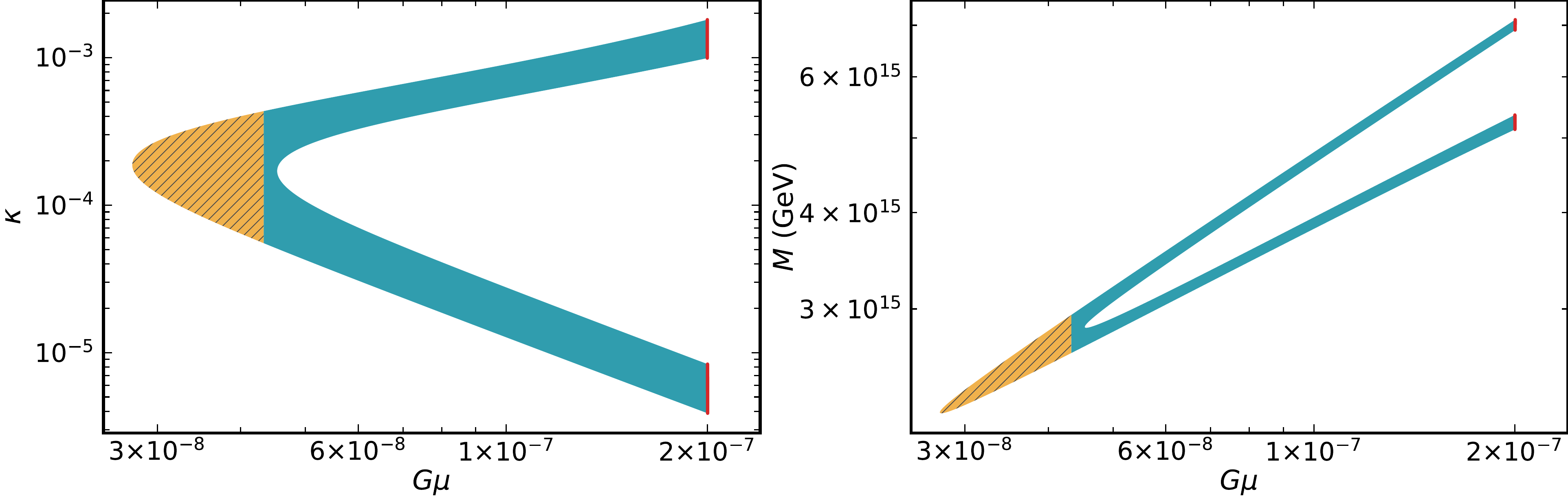}
    \caption{Behavior of $\kappa$ (left) and the symmetry breaking scale $M$ (right) with respect to the dimensionless parameter $G\mu$. The blue shaded region represents the Planck2018 2-$\sigma$ bounds, while the yellow hatched region is excluded by the NANOGrav 15-year data bound on $G\mu$. The red cutoff on the right corresponds to the third advanced LVK bound on $G\mu$.}
    \label{if_pr_1}
\end{figure*}

\begin{figure*}[ht!]
\centering\includegraphics[width=\textwidth]{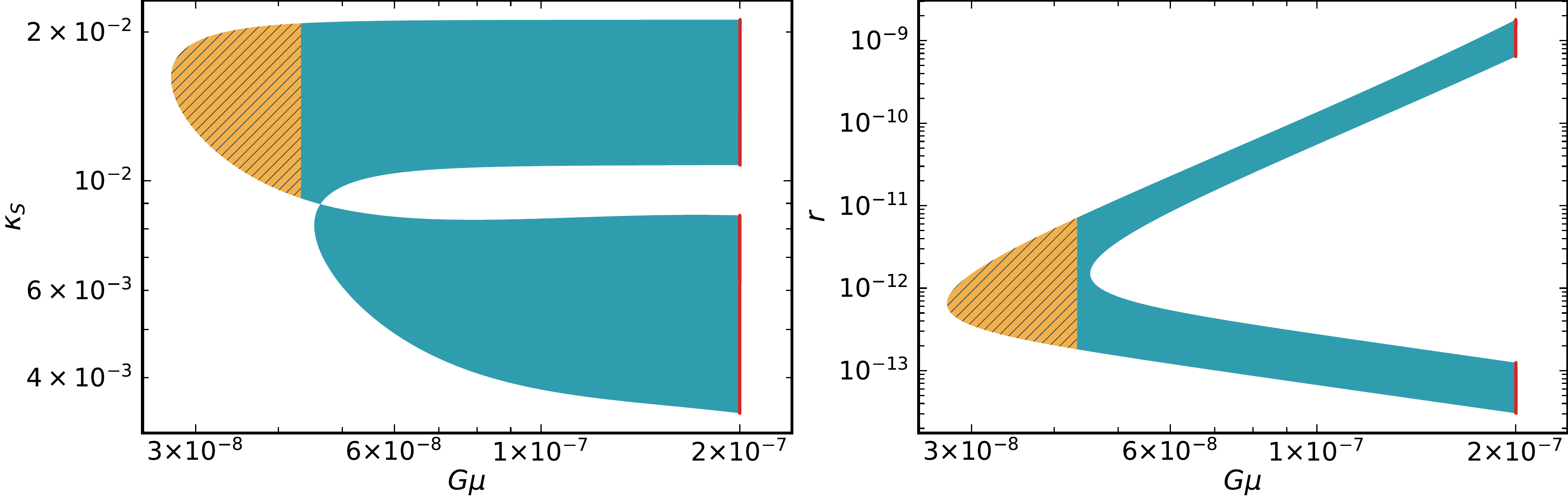}
    \caption{Behavior of the non-minimal coupling $\kappa_S$ (left) and the tensor-to-scalar ratio $r$ (right) with respect to the dimensionless parameter $G\mu$. The blue shaded region represents the Planck2018 2-$\sigma$ bounds, while the yellow hatched region is excluded by the NANOGrav 15-year data bound on $G\mu$. The red cutoff on the right corresponds to the third advanced LVK bound on $G\mu$.}
    \label{if_pr_2}
\end{figure*}

\begin{figure*}[ht!]
\centering\includegraphics[width=\textwidth]{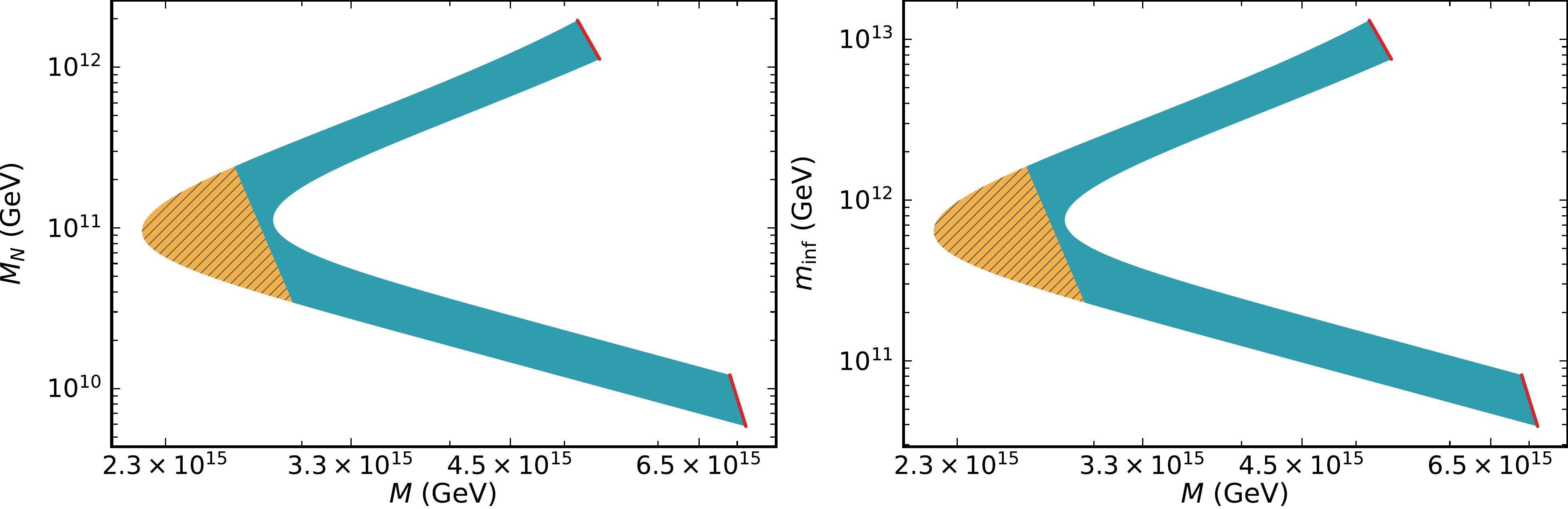}
    \caption{Behavior of the RHN mass $M_N$ (left) and the inflaton mass $m_{\text{inf}}$ (right) with respect to the symmetry-breaking scale $M$. The blue shaded region represents the Planck2018 2-$\sigma$ bounds, while the yellow hatched region is excluded by the NANOGrav 15-year data bound on $G\mu$. The red cutoff on the right corresponds to the third advanced LVK bound on $G\mu$.}
    \label{if_pr_3}
\end{figure*}

\section{Numerical analysics}\label{Nans}
In this section we analyse the implications of the model and discuss its predictions regarding the various cosmological observables. We pay particular attention to the parametric space which is consistent with NANOGrav 15 year data. Before presenting numerical predictions we briefly review the basic results of the slow-roll assumption. The prediction for the various inflationary parameters are calculated using the standard slow-roll parameters,
\begin{align}
\epsilon &= \frac{1}{4}\left( \frac{m_P}{M}\right)^2
\left( \frac{V'}{V}\right)^2, \,\,\,
\eta = \frac{1}{2}\left( \frac{m_P}{M}\right)^2
\left( \frac{V''}{V} \right),\nonumber\\ 
\xi^2 &= \frac{1}{4}\left( \frac{m_P}{M}\right)^4
\left( \frac{V' V'''}{V^2}\right),
\label{slowroll}
\end{align}
where, prime denotes the derivative with respect to $x$. The scalar spectral index $n_s$, the tensor-to-scalar ratio $r$, and the running of the scalar spectral index $\alpha_{s}\equiv dn_s / d \ln k$, in the slow-roll approximation are given by,
\begin{align} \label{nsr}
n_s&\simeq 1+2\,\eta-6\,\epsilon, \\
r &\simeq 16\,\epsilon, \\
\alpha_{s} &\simeq 16\,\epsilon\,\eta
-24\,\epsilon^2 - 2\,\xi^2.
\end{align}
The observational constraint on scalar spectral index $n_s$ from Planck 2018 data in the base $\Lambda$CDM model at 68 \% CL is given as \cite{Planck:2018jri} ,
\begin{equation}
    n_s = 0.9665 \pm 0.0038.
\end{equation}
The amplitude of the scalar power spectrum is given by,
\begin{align}
A_{s}(k_0) = \frac{1}{24\,\pi^2\,\epsilon(x_0)}
\left( \frac{V(x_0)}{m_P^4}\right),  \label{curv}
\end{align}
which at the pivot scale $k_0 = 0.05\, \rm{Mpc}^{-1}$ is given by $A_{s}(k_0) = 2.137 \times 10^{-9}$, as measured by Planck 2018 \cite{Planck:2018jri}.
The last number of e-folds, $N_0$, is given by,
\begin{align}\label{Ngen}
N_0 = 2\left( \frac{M}{m_P}\right) ^{2}\int_{x_e}^{x_{0}}\left( \frac{V}{%
	V'}\right) dx,
\end{align}
where $x_0 \equiv x(k_0)$ and
$x_e$ are the field values at the pivot scale $k_0$ and at the end of inflation, respectively. The value of $x_e$ is determined by the breakdown of the slow-roll approximation. In our model, there are seven independent key parameters:  $\kappa$, $M$, $am_{3/2}$, $\kappa_{S}$, $\kappa_{SS}$, $x_0$, and $x_e$. To simplify the analysis, we set $\kappa_{S} = \kappa_{SS}$, and fix $am_{3/2}$ at $10$ TeV for convenience. The remaining parameters are subject to several essential constraints:

\begin{itemize}
    \item The amplitude of the scalar power spectrum, denoted as $A_s(k_0)$, with a specific value of $2.137\times 10^{-9}$ (as given in Eq. \eqref{curv})
    \item  The end of inflation, determined by the waterfall mechanism, with the condition that $x_e=1$.
    \item The number of e-folds, represented as $N_0$ in Eq. \eqref{Ngen}, is determined by $T_R$ as defined in Eq. \eqref{efolds}. In this scenario, we specifically set $T_R = 10^{9}$ GeV to avoid the gravitino problem.
    \item The observed value of the baryon-to-photon $n_B/n_{\gamma}$, which takes the specific value of $6.12\times 10^{-10}$ (as given in Eq. \eqref{bphr}).
\end{itemize}
These constraints play a crucial role in determining the predictions of the model. With these constraints imposed, the parameters $\kappa$, $\kappa_{S} = \kappa_{SS}$, and $M$ are varied. The results of our numerical calculations are presented in \Cref{if_pr_1} - \ref{if_pr_3}, where the behavior of various parameters is shown across different planes. The blue region represents the allowed parametric space between $4.33 \times 10^{-8} \lesssim G\mu  \lesssim  2 \times 10^{-7}$, which is consistent with the Planck 2018 2$\sigma$ bounds on the scalar spectral index $n_s$, as well as the NANOGrav 15-year data bound and the third advanced LIGO/Virgo/KAGRA (LVK) bound on the dimensionless string tension parameter $G\mu$. The yellow hatched region on the left is excluded by the NANOGrav 2$\sigma$ bounds, while the red cutoff on the right represents the third advanced LVK bound on $G\mu$ which is slightly stronger than the CMB bound 
\begin{equation}
    G\mu \lesssim 1.3 \times 10^{-7}.
\end{equation}
 
\Cref{if_pr_1} depicts the variation of the dimensionless parameter $\kappa$ and the symmetry breaking scale $M$ with $G\mu$, while \Cref{if_pr_2} illustrates the variation of the non-minimal coupling $\kappa_S$ and the tensor-to-scalar ratio $r$ with $G\mu$. It is evident that the parametric space consistent with the cosmic string bounds yields small tensor modes, which are unlikely to be observable in any of the forthcoming CMB experiments. This can be understood from the following relationship between $r$, $\kappa$ and $M$
\begin{equation}
    r \simeq \left( \frac{2 \kappa^2}{3 \pi^2 A_s (k_0)}\right) \left( \frac{M}{m_P} \right)^4.
\end{equation}
Since both $M$ and $\kappa$ are small, we expect tiny values of the tensor-to-scalar ratio. Specifically, for $M \simeq 5 \times 10^{15}$ GeV and $\kappa \simeq 2 \times 10^{-3}$, the above equation yields $r \sim 2 \times 10^{-9}$, which closely aligns with the more accurate values obtained in our numerical results. Furthermore, the scalar spectral index $n_s$ can be estimated from Eq. \eqref{nsr} as
\begin{align}
    n_s &\simeq 1 + \left( \frac{m_P}{M} \right)^2 \left( 6 \gamma_S x_0^2 \left( \frac{M}{m_P} \right)^4 - 2 \kappa_S \left( \frac{M}{m_P} \right)^2 \right. \nonumber \\ 
    & + \left.  \frac{\mathcal{N} \kappa^2}{8 \pi^2} F^{\prime \prime} (x_0)  +  \frac{\gamma_h^3 \kappa^2 F^{\prime \prime}(\sqrt{\gamma_h} x_0)}{16 \pi^2} \right).
\end{align}
For $\kappa \simeq 10^{-6}$, the radiative corrections are suppressed while the SUGRA corrections dominate. With $x_0 \simeq 1$, $\kappa_S \simeq 0.021$, and $M \simeq 7 \times 10^{15}$ GeV, the above equation gives $n_s \simeq 0.9578$, which agrees well with the values obtained in our numerical results. 

\Cref{if_pr_3} displays the variation of the right-handed neutrino mass and inflaton mass with the gauge symmetry breaking scale $M$. These ranges of $M_N$ and $m_{\text{inf}}$ are consistent with leptogenesis and satisfy the constraints from the observed baryon asymmetry of the universe. The bound $m_{\text{inf}} > 2 M_N$ is satisfied throughout the entire parametric space, and since $M_N \gg T_R$, the out-of-equilibrium condition necessary for successful implementation of leptogenesis is automatically met.

The gauge-breaking scale $M$ is related to the cosmic string tension parameter $\mu$. In our investigation, we focus on models where the breaking of Abelian symmetry is associated with the vacuum expectation value (vev) of multiplets responsible for the masses of Right-Handed Neutrinos (RHNs). The string tension is approximately given by:
\begin{align}\label{mucs}
\mu_s &= 2\pi M^2 \epsilon(\beta), \notag\\	\epsilon(\beta)&=\frac{2.4}{\log(2/\beta)} \text{  for  } \beta=\frac{\kappa^2}{2g^2}<10^{-2},
\end{align}
where $g = g_{1Y}(M) \simeq 0.7$ for MSSM. When a simple group undergoes a breakdown into a subgroup containing an Abelian factor, it leads to the formation of monopoles. However, to prevent the overclosure of the universe, inflationary processes must eliminate these monopoles. Subsequently, after the remaining Abelian symmetry is broken, cosmic strings emerge. When the scales of monopoles and cosmic strings are close, Schwinger nucleation occurs, generating monopole-antimonopole pairs on the string, initiating its decay. The decay process depends on the ratio of the monopole and string formation scales, defined by:
\begin{align}
\kappa_m = \frac{M_{m}^2}{\mu_{s}}\sim \frac{8\pi}{g_m^2} \left( \frac{v_m}{M} \right)^2,   \label{eq:alpha} 
\end{align}
where $M_{m}$ represents the monopole mass, $v_m$ denotes its formation scale, and $g_m = g_4(v_m) = g_3(v_m)$ stands for the gauge coupling of the symmetry responsible for monopole generation. Assuming $g_m \sim 1 $, the observed NANOGrav signal can be explained by metastable cosmic strings, with $\sqrt{\kappa_m} \sim 8$, implying $v_m \sim 2 M$.

Cosmic strings dissipate energy through gravitational radiation. For stable strings, they emit gravitational waves until their entire energy is converted, leading to the disappearance of the string network. While metastable strings initially behave like stable ones, they eventually decay due to monopole-antimonopole pair production, with a decay time given by
\begin{align}
t_s=\Gamma^{-1/2}_d, \;\;\;\;\; \Gamma_d=\frac{\mu}{2\pi} e^{-\pi\kappa_m}.     
\end{align}
The exponential suppression in $\Gamma_d$ implies that metastable cosmic string networks mimic stable networks when $\kappa_m^{1/2}\gg 10$.

In summary, for the non-minimal coupling $0.0034 \lesssim \kappa_S \lesssim 0.021$, and the scalar spectral index fixed at the 2$\sigma$ bounds of Planck 2018 data, we obtain $3.8 \times 10^{-6} \lesssim \kappa \lesssim 1.8 \times 10^{-3}$, symmetry breaking scale $2.2 \times 10^{15} \text{ GeV} \lesssim M \lesssim 7.1 \times 10^{15} \text{ GeV}$, inflaton mass $3.9 \times 10^{10} \text{ GeV} \lesssim m_{\text{inf}} \lesssim 1.3 \times 10^{13} \text{ GeV}$, and right-handed neutrino mass $5.8 \times 10^{9} \text{ GeV} \lesssim M_N \lesssim 2 \times 10^{12} \text{ GeV}$ with $2.8 \times 10^{-4} \lesssim \alpha_1 \lesssim 1.8 \times 10^{-1}$. Intriguingly, the entire region of the parameter space obtained in our results, with $4.33 \times 10^{-8} \lesssim G\mu  \lesssim  2 \times 10^{-7}$ for $\sqrt{\kappa_m} \simeq 8$, lies across a broad spectrum of frequencies that will be fully probed by several gravitational wave observatories (Ref. \cite{NANOGrav:2023hvm, Ferdman:2010xq, LISA:2017pwj, Hu:2017mde, TianQin:2015yph, Corbin:2005ny, Seto:2001qf, Punturo:2010zz, LIGOScientific:2016wof, AEDGE:2019nxb}). \footnote{A significant amount of research has been conducted in the realm of stochastic gravitational waves in recent years. For detailed see references \cite{Ahmed:2023pjl, Ahmed:2023rky, Afzal:2023cyp, King:2023wkm, Lazarides:2023rqf, Lazarides:2023ksx, Buchmuller:2023aus, Antusch:2023zjk, Fu:2023mdu, Vagnozzi:2023lwo, Vagnozzi:2020gtf, Buchmuller:2021dtt, Buchmuller:2021mbb, Masoud:2021prr, Ahmed:2022rwy, Afzal:2022vjx, Pallis:2024mip}.}

\section{The \texorpdfstring{$G_{421} \subset G_{422}$}{421 to 422} Embedding and Proton Decay} \label{sec2}%
\begin{figure}[t!]
    \centering
    \includegraphics[width=0.7\linewidth]{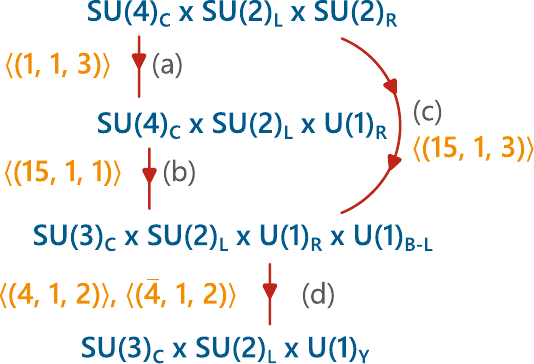}
    \caption{\label{422}Breaking patterns of the Pati-Salam gauge symmetry $G_{422}$ down to the standard model gauge symmetry $G_{\text{SM}}$.}
\end{figure}

	\begin{table}[t!]
    \caption{\label{422content} Decomposition of Pati-Salam content into $G_{421}$ representations and their $U(1)_{R'}$ charges.}
    \begin{ruledtabular}
        \begin{tabular}{ccl}
         $G_{422}$&$q\bigl(U(1)_{R'}\bigr)$&
        \qquad$  G_{421}$ \\
        \hline
         $F_i(4,2,1)$  &$1/2$& $F_i(4,2,0)$ \\
         $F^c_i(\bar{4},1,2)$  &$1/2$&$F^c_u(\bar{4},1,-1/2)+F^c_d(\bar{4},1,1/2)$ \\
        \hline
         $\overline{H}^c(4,1,2)$ &$0$&$ H(4,1,1/2)+H'(4,1,-1/2)$ \\
         $H^c(\bar{4},1,2)$ & $0$ &$\overline{H}(\bar{4},1,-1/2)+\overline{H}'(\bar{4},1,1/2)$ \\
         $h(1,2,2)$& $0$ & $h_u(1,2,1/2)+h_d(1,2,-1/2)$ \\
         $G(6,1,1)$ & $1$ & $G(6,1,0)$\\
         $S(1,1,1)$ & $1$ & $S(1,1,0)$
        \end{tabular}
    \end{ruledtabular}
\end{table}

\begin{table*}[t!]
\caption{\label{addcontent}Decomposition of new superfields, added in $G_{421}$ content, into $G_{3211}$ and $G_{321}$, and their $U(1)_{R'}$ charges. }
    \begin{ruledtabular}
        \begin{tabular}{cccllc}
            &$G_{421}$&
            $q\bigl(U(1)_{R'}\bigr)$&
            \qquad$  G_{3211}$&
            \qquad$  G_{321}$&\\
            \hline
            &${H}' \left( 4, 1, -1/2 \right)$ &  $0$    &         $d_H\left( 3, 1, -1/2, 1/3 \right)$  & $ {d}_H\left( 3, 1, -1/3 \right)$ &\\
			&& & $ {e}_H\left(1, 1, -1/2, -1 \right)$ & $ {e}_H\left(1, 1, -1 \right)$& \\
			&$\overline{H}' \left( \bar{4}, 1, 1/2 \right)$ &  $0$    &         $\bar{d}_H\left( \bar{3}, 1, 1/2, -1/3 \right)$  & $\bar{d}_H\left( \bar{3}, 1, 1/3 \right)$ &\\
			&& & $\bar{e}_H\left(1, 1, 1/2, 1 \right)$ & $\bar{e}_H\left(1, 1, 1 \right)$& \\
  & $G(6,1,0)$& $1$ &$g_a(3,1,0,-2/3)$& $g_a(3,1,-1/3)$ & \\
   &&&$g^c_a(\bar{3},1,0,2/3)$& $g^c_a(\bar{3},1,1/3)$ &\\
            \end{tabular}
    \end{ruledtabular}
\end{table*}

The subgroup $G_{421} \equiv SU(4)_C \times SU(2)_L \times U(1)_{R}$ can be embedded within the larger Pati-Salam  group $G_{422} \equiv SU(4)_C \times SU(2)_L \times SU(2)_R$. The breaking patterns of the $G_{422}$  symmetry down to the standard model gauge symmetry are illustrated in \cref{422}. In the first breaking pattern, $G_{422}$ breaks to $G_{421}$ via $\langle (1,1,3)\rangle$ (a), then $G_{421}$ subsequently breaks to $G_{3211}$ via $\langle (15,1,1)\rangle$ (b). In the alternative breaking pattern, $G_{422}$ directly breaks to $G_{3211}$ via $\langle (15,1,3)\rangle$ (c). Finally, in both breaking patterns, the group $G_{3211}$ breaks to the SM gauge group $G_{\text{SM}}$ via $\langle (4, 1, 2)\rangle$ and $\langle (\bar{4}, 1, 2)\rangle$ (d). In this context, we consider the breaking pattern for $G_{422}$  symmetry involving pathways (a) and (b).

The decomposition of $G_{422}$ representations employed in this model under the $G_{421}$ gauge group is provided in \cref{422content}, along with their $U(1)_R'$ charge assignments. The MSSM matter content and the right-handed neutrino (RHN) superfields reside in the $F_i(4,2,1)$ and $F^c_i(\bar{4},1,2)$ representation of $G_{422}$, while the Higgs sector resides in $\overline{H}^c(4,1,2)$, $H^c(\bar{4},1,2)$ and $h(1,2,2)$ representations.
To embed $G_{421}$ in $G_{422}$, we need to add superfields $H'$ and $\overline{H}'$ in $G_{421}$ content in \cref{content}. Decomposition of $H'$ and $\overline{H}'$ under $G_{3211}$ and $G_{\text{SM}}$ along with their $U(1)_{R'}$ charges are given in \cref{addcontent}. The addition of these superfields results in the appearance of the following non-renormalizable terms in the superpotential
\begin{eqnarray}
    W &\supset & \frac{ H' H}{m_P}( \gamma_3 F F + \gamma_4 F^c_u F^c_d)
\nonumber \\
&+&\frac{ \overline{H}' \overline{H} }{m_P} ( \gamma_5 F F + \gamma_6 F^c_u F^c_d),
\end{eqnarray}
which contribute to the proton decay. Thus, the content of our $G_{421}$ model explains the predictions of proton decay in the $G_{422}$ model \cite{Lazarides:2020bgy}. Integrating out the color triplets we effectively provide proton decay operators. For color triplets $(d_H, \bar{d}_H)$ to acquire mass, $G_{422}$  and $G_{421}$ must contain a sextet superfield $G$ such that,
\begin{eqnarray}
    W &\supset & \lambda_{T}\ G H^c H^c + \lambda_{\bar{T}}\ G \overline{H}^c \overline{H}^c, \\
    &\supset & \lambda_{T}\ G \overline{H}' \overline{H} + \lambda_{\bar{T}}\ G H H',
\end{eqnarray}
where $\lambda_{T}$ and $\lambda_{\bar{T}}$ are dimensionless couplings. Thus, masses of color  triplets $d_H$ and $\bar{d}_H$ become,
\begin{eqnarray}
    M_{d_H}=\lambda_{\bar{T}} M,\qquad M_{\bar{d}_H}=\lambda_{T} M.
\end{eqnarray}
We assume $\lambda_{T}\sim \lambda_{\bar{T}}$ such that $M_{d_H}=M_{\bar{d}_H}=M_T$. 
Decomposition of $G$ and its $U(1)_{R'}$ charge is given in \cref{addcontent}.
    \begin{figure}[t!]\centering
\subfloat[$ QQ(U^cE^c)^\dagger$ ]{\includegraphics[width=0.17\textwidth]{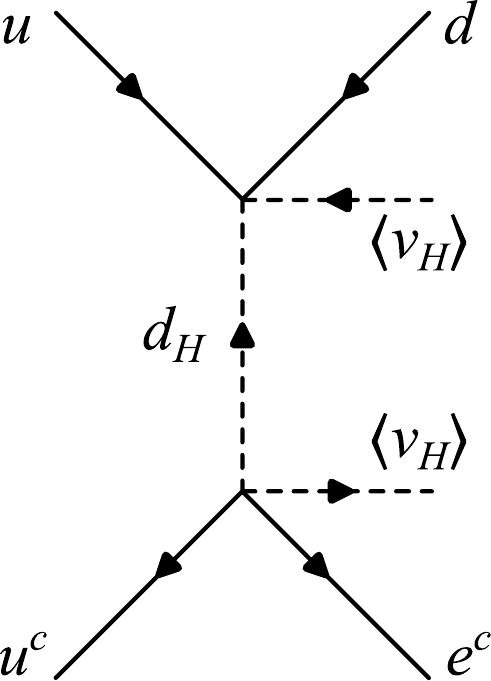}}\qquad
    \subfloat[$ QL(U^cD^c)^\dagger$ ]{\includegraphics[width=0.20\textwidth]{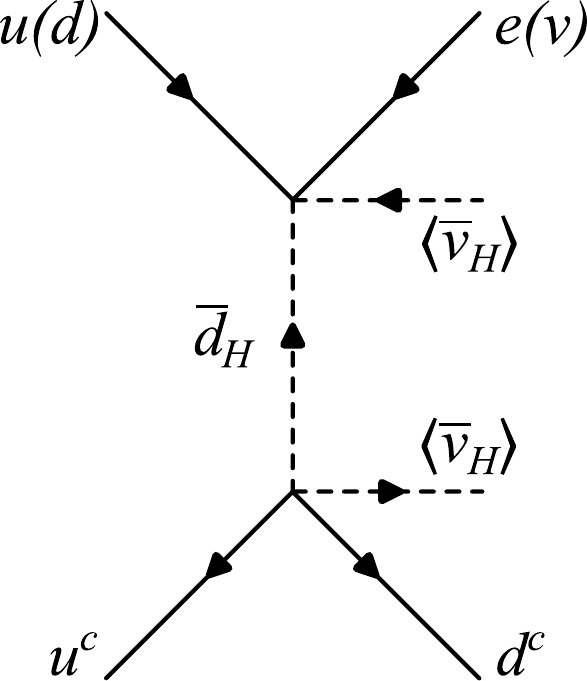}}
    \caption{\label{LLRR}Four fermion LLRR chirality non-flipping dimension six proton decay operators mediated via color triplets $d_H$, $\bar{d}_H$.}
    \end{figure}
\begin{figure}[ht!]
    \centering
    \includegraphics[width=1.0\linewidth]{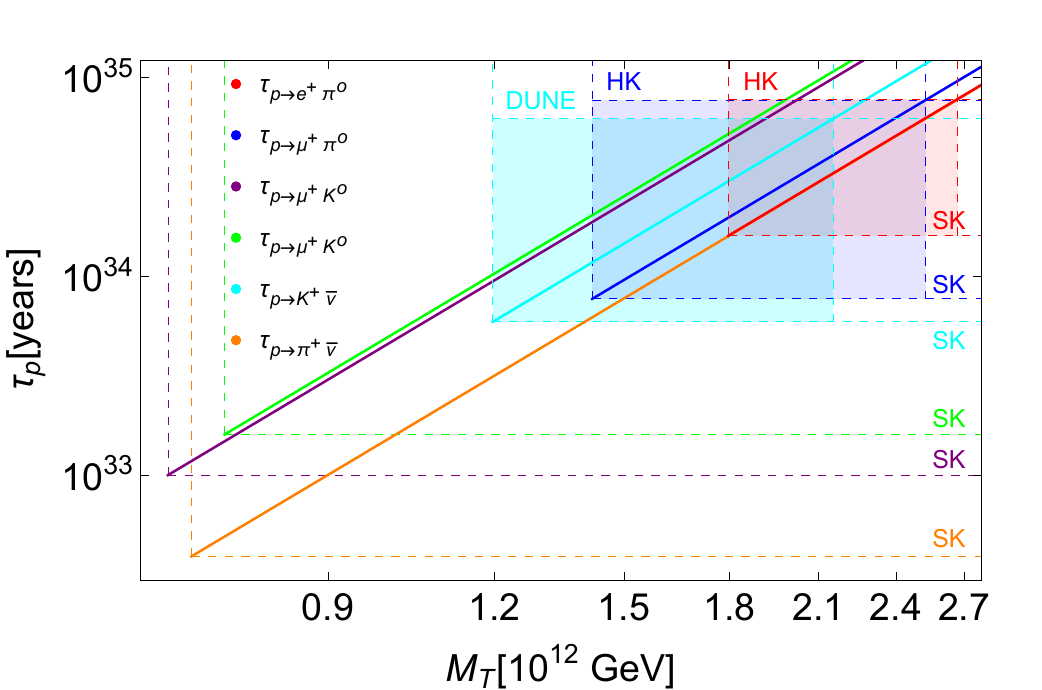}
    \caption{\label{pdecay}Predictions of proton lifetime for various decay channels, potentially observable in the next-generation of experiments such as Hyper Kamiokande and DUNE.}
\end{figure}

Proton decay predictions in $G_{421}$ are the same as already discussed in $G_{422}$  in \cite{Lazarides:2020bgy}. Here we briefly recap the decay rates and explore the parameter range for the predictions observable in next-generation experiments. Dominant contribution in the decay rate of proton comes from chirality non-flipping LLRR dimension six four fermion operators, shown in \cref{LLRR}.
To explore the parameter range that predicts observable proton decay we assumed $\gamma_{k}=\gamma \sim \mathcal{O}(1)$, for $k = 3,4,5,6$. Decay rates for charged and neutral lepton channels become,
 \begin{eqnarray}
    \Gamma_{p\rightarrow \pi^0 l^+_i}&=&k_{\pi}\left|\frac{M^2}{m_P^2}\frac{\gamma^2}{M_T^2}T_{\pi^0 l^+}\right|^2\left(A_{S1}^2+A_{S2}^2\right),\\
    \Gamma_{p\rightarrow K^0 l^+_i}&=&k_{K}\left|\frac{M^2}{m_P^2}\frac{\gamma^2}{M_T^2}T_{K^0 l^+}\right|^2\left(A_{S1}^2+A_{S2}^2\right),\\
    \Gamma_{p\rightarrow \pi^+ \bar{\nu}_i}&=&k_{\pi}\left|\frac{M^2}{m_P^2}\frac{\gamma^2}{M_T^2}T_{\pi^+ \bar{\nu}}A_{S2}\right|^2 ,\qquad\\
    \Gamma_{p\rightarrow K^+ \bar{\nu}_i}&=&k_{K}\left|\frac{M^2}{m_P^2}\frac{\gamma^2}{M_T^2}A_{S2}\right|^2\left|T'_{K^+ \bar{\nu}}+T''_{K^+ \bar{\nu}}\right|^2, \qquad
\end{eqnarray}   
where,
\begin{eqnarray}
    k_{\pi/K}=\frac{m_p A_L^2}{32\pi}\left(1-\frac{m_{\pi/K}^2}{m_p^2}\right)^2.
\end{eqnarray}
The decay rates for different proton decay channels are graphed in \cref{pdecay}, showcasing the lifetime of the proton against $M_T$. In \cref{pdecay}, the notation `SK' denotes the experimental limit on the proton lifetime across different decay channels, as determined by Super Kamiokande \cite{ParticleDataGroup:2018ovx}. On the other hand, `HK' and `DUNE' signify the anticipated sensitivities of forthcoming experiments: Hyper Kamiokande \cite{Hyper-Kamiokande:2018ofw} and DUNE \cite{DUNE:2020ypp,DUNE:2015lol}, respectively, in detecting various proton decay channels. We note that when $M_T$ is $\mathcal{O}(10^{12}\ \text{GeV})$, the model predicts observable proton decay rates, detectable by experiments such as Hyper Kamiokande and DUNE. This trend is consistent across the entire range of $M$ investigated in \cref{Nans}.
\section{Summary}\label{sec8}

In conclusion, we have investigated the $\mu$-hybrid inflation model within the context of supersymmetric $SU(4)_C \times SU(2)_L \times U(1)_R$ gauge symmetry, which spontaneously breaks to the subgroup $SU(3)_C \times SU(2)_L \times U(1)_R \times U(1)_{B-L}$ before inflation. Subsequent breaking of $U(1)_{R} \times U(1)_{B-L} \rightarrow U(1)_Y$ after inflation avoids the primordial magnetic monopole problem and leads to the formation of a metastable cosmic string network whose decay through Schwinger nucleation of monopole-antimonopole pairs generates a stochastic gravitational wave background (SGWB) consistent with current PTA data. The predictions of the scalar spectral index $n_s$ and tensor-to-scalar ratio $r$ turn out to be perfectly consistent with Planck 2018 bounds. Additionally, we have briefly explored the integration of $G_{421}$ into the Pati-Salam symmetry through the inclusion of additional superfields such as $H^{\prime} (4, 1, -1/2)$, $\overline{H}^{\prime} (\bar{4}, 1, 1/2)$, and a sextet $G (6, 1, 0)$. This minimal extension allows for the proton decay predictions of the $G_{422}$ model within the context of the $G_{421}$ model and provides a natural explanation for the intermediate-scale mass of color triplets. These triplets play a pivotal role in predicting observable proton decay rates, making them particularly interesting for upcoming experiments like Hyper Kamiokande and DUNE.


\FloatBarrier
\nocite{*}
\bibliographystyle{apsrev4-1}
\bibliography{bibliography}
\end{document}

%% file: authors.tex
\author{Waqas Ahmed
\orcidlink{0000-0001-8136-9958}}
\email{waqasmit@hbpu.edu.cn}
\affiliation{Center for Fundamental Physics and School of Mathematics and Physics, Hubei Polytechnic University, Huangshi 435003, China}
\affiliation{University of Europe for Applied Sciences Innovation Hub, Konrad-Zuse-Ring 11 · 14469 Potsdam,
Germany.}
\author{Maria Mehmood 
\orcidlink{0000-0002-3792-8561}}
\email{mehmood.maria786@gmail.com}
\affiliation{Department of Physics, Quaid-i-Azam University, Islamabad, 45320, Pakistan}
\author{Mansoor Ur Rehman
\orcidlink{0000-0002-1780-1571}}
\email{mansoor@qau.edu.pk}
\affiliation{Department of Physics, Quaid-i-Azam University, Islamabad, 45320, Pakistan}
\author{Umer Zubair
\orcidlink{0000-0003-1671-8722}}
\email{umer@udel.edu}
\affiliation{Department of Physics and Astronomy, University of Delaware, Newark, DE 19716, USA}

%% file: SGWB_421_draft_V1.bbl
\begin{thebibliography}{91}%
\makeatletter
\providecommand \@ifxundefined [1]{%
 \@ifx{#1\undefined}
}%
\providecommand \@ifnum [1]{%
 \ifnum #1\expandafter \@firstoftwo
 \else \expandafter \@secondoftwo
 \fi
}%
\providecommand \@ifx [1]{%
 \ifx #1\expandafter \@firstoftwo
 \else \expandafter \@secondoftwo
 \fi
}%
\providecommand \natexlab [1]{#1}%
\providecommand \enquote  [1]{``#1''}%
\providecommand \bibnamefont  [1]{#1}%
\providecommand \bibfnamefont [1]{#1}%
\providecommand \citenamefont [1]{#1}%
\providecommand \href@noop [0]{\@secondoftwo}%
\providecommand \href [0]{\begingroup \@sanitize@url \@href}%
\providecommand \@href[1]{\@@startlink{#1}\@@href}%
\providecommand \@@href[1]{\endgroup#1\@@endlink}%
\providecommand \@sanitize@url [0]{\catcode `\\12\catcode `\$12\catcode
  `\&12\catcode `\#12\catcode `\^12\catcode `\_12\catcode `\%12\relax}%
\providecommand \@@startlink[1]{}%
\providecommand \@@endlink[0]{}%
\providecommand \url  [0]{\begingroup\@sanitize@url \@url }%
\providecommand \@url [1]{\endgroup\@href {#1}{\urlprefix }}%
\providecommand \urlprefix  [0]{URL }%
\providecommand \Eprint [0]{\href }%
\providecommand \doibase [0]{http://dx.doi.org/}%
\providecommand \selectlanguage [0]{\@gobble}%
\providecommand \bibinfo  [0]{\@secondoftwo}%
\providecommand \bibfield  [0]{\@secondoftwo}%
\providecommand \translation [1]{[#1]}%
\providecommand \BibitemOpen [0]{}%
\providecommand \bibitemStop [0]{}%
\providecommand \bibitemNoStop [0]{.\EOS\space}%
\providecommand \EOS [0]{\spacefactor3000\relax}%
\providecommand \BibitemShut  [1]{\csname bibitem#1\endcsname}%
\let\auto@bib@innerbib\@empty
\bibitem [{\citenamefont {Agazie}\ \emph
  {et~al.}(2023{\natexlab{a}})\citenamefont {Agazie} \emph
  {et~al.}}]{NANOGrav:2023gor}%
  \BibitemOpen
  \bibfield  {author} {\bibinfo {author} {\bibfnamefont {G.}~\bibnamefont
  {Agazie}} \emph {et~al.} (\bibinfo {collaboration} {NANOGrav}),\ }\href
  {\doibase 10.3847/2041-8213/acdac6} {\bibfield  {journal} {\bibinfo
  {journal} {Astrophys. J. Lett.}\ }\textbf {\bibinfo {volume} {951}},\
  \bibinfo {pages} {L8} (\bibinfo {year} {2023}{\natexlab{a}})},\ \Eprint
  {http://arxiv.org/abs/2306.16213} {arXiv:2306.16213 [astro-ph.HE]}
  \BibitemShut {NoStop}%
\bibitem [{\citenamefont {Antoniadis}\ \emph
  {et~al.}(2023{\natexlab{a}})\citenamefont {Antoniadis} \emph
  {et~al.}}]{EPTA:2023fyk}%
  \BibitemOpen
  \bibfield  {author} {\bibinfo {author} {\bibfnamefont {J.}~\bibnamefont
  {Antoniadis}} \emph {et~al.} (\bibinfo {collaboration} {EPTA}),\ }\href@noop
  {} {\  (\bibinfo {year} {2023}{\natexlab{a}})},\ \Eprint
  {http://arxiv.org/abs/2306.16214} {arXiv:2306.16214 [astro-ph.HE]}
  \BibitemShut {NoStop}%
\bibitem [{\citenamefont {Reardon}\ \emph {et~al.}(2023)\citenamefont {Reardon}
  \emph {et~al.}}]{Reardon:2023gzh}%
  \BibitemOpen
  \bibfield  {author} {\bibinfo {author} {\bibfnamefont {D.~J.}\ \bibnamefont
  {Reardon}} \emph {et~al.},\ }\href {\doibase 10.3847/2041-8213/acdd02}
  {\bibfield  {journal} {\bibinfo  {journal} {Astrophys. J. Lett.}\ }\textbf
  {\bibinfo {volume} {951}},\ \bibinfo {pages} {L6} (\bibinfo {year} {2023})},\
  \Eprint {http://arxiv.org/abs/2306.16215} {arXiv:2306.16215 [astro-ph.HE]}
  \BibitemShut {NoStop}%
\bibitem [{\citenamefont {Xu}\ \emph {et~al.}(2023)\citenamefont {Xu} \emph
  {et~al.}}]{Xu:2023wog}%
  \BibitemOpen
  \bibfield  {author} {\bibinfo {author} {\bibfnamefont {H.}~\bibnamefont {Xu}}
  \emph {et~al.},\ }\href {\doibase 10.1088/1674-4527/acdfa5} {\bibfield
  {journal} {\bibinfo  {journal} {Res. Astron. Astrophys.}\ }\textbf {\bibinfo
  {volume} {23}},\ \bibinfo {pages} {075024} (\bibinfo {year} {2023})},\
  \Eprint {http://arxiv.org/abs/2306.16216} {arXiv:2306.16216 [astro-ph.HE]}
  \BibitemShut {NoStop}%
\bibitem [{\citenamefont {Agazie}\ \emph
  {et~al.}(2023{\natexlab{b}})\citenamefont {Agazie} \emph
  {et~al.}}]{NANOGrav:2023hfp}%
  \BibitemOpen
  \bibfield  {author} {\bibinfo {author} {\bibfnamefont {G.}~\bibnamefont
  {Agazie}} \emph {et~al.} (\bibinfo {collaboration} {NANOGrav}),\ }\href
  {\doibase 10.3847/2041-8213/ace18b} {\bibfield  {journal} {\bibinfo
  {journal} {Astrophys. J. Lett.}\ }\textbf {\bibinfo {volume} {952}},\
  \bibinfo {pages} {L37} (\bibinfo {year} {2023}{\natexlab{b}})},\ \Eprint
  {http://arxiv.org/abs/2306.16220} {arXiv:2306.16220 [astro-ph.HE]}
  \BibitemShut {NoStop}%
\bibitem [{\citenamefont {Afzal}\ \emph
  {et~al.}(2023{\natexlab{a}})\citenamefont {Afzal} \emph
  {et~al.}}]{NANOGrav:2023hvm}%
  \BibitemOpen
  \bibfield  {author} {\bibinfo {author} {\bibfnamefont {A.}~\bibnamefont
  {Afzal}} \emph {et~al.} (\bibinfo {collaboration} {NANOGrav}),\ }\href
  {\doibase 10.3847/2041-8213/acdc91} {\bibfield  {journal} {\bibinfo
  {journal} {Astrophys. J. Lett.}\ }\textbf {\bibinfo {volume} {951}},\
  \bibinfo {pages} {L11} (\bibinfo {year} {2023}{\natexlab{a}})},\ \Eprint
  {http://arxiv.org/abs/2306.16219} {arXiv:2306.16219 [astro-ph.HE]}
  \BibitemShut {NoStop}%
\bibitem [{\citenamefont {Pati}\ and\ \citenamefont
  {Salam}(1974)}]{Pati:1974yy}%
  \BibitemOpen
  \bibfield  {author} {\bibinfo {author} {\bibfnamefont {J.~C.}\ \bibnamefont
  {Pati}}\ and\ \bibinfo {author} {\bibfnamefont {A.}~\bibnamefont {Salam}},\
  }\href {\doibase 10.1103/PhysRevD.10.275} {\bibfield  {journal} {\bibinfo
  {journal} {Phys. Rev. D}\ }\textbf {\bibinfo {volume} {10}},\ \bibinfo
  {pages} {275} (\bibinfo {year} {1974})},\ \bibinfo {note} {[Erratum:
  Phys.Rev.D 11, 703--703 (1975)]}\BibitemShut {NoStop}%
\bibitem [{\citenamefont {Georgi}(1975)}]{Georgi:1974my}%
  \BibitemOpen
  \bibfield  {author} {\bibinfo {author} {\bibfnamefont {H.}~\bibnamefont
  {Georgi}},\ }\href {\doibase 10.1063/1.2947450} {\bibfield  {journal}
  {\bibinfo  {journal} {AIP Conf. Proc.}\ }\textbf {\bibinfo {volume} {23}},\
  \bibinfo {pages} {575} (\bibinfo {year} {1975})}\BibitemShut {NoStop}%
\bibitem [{\citenamefont {Gell-Mann}\ \emph {et~al.}(1979)\citenamefont
  {Gell-Mann}, \citenamefont {Ramond},\ and\ \citenamefont
  {Slansky}}]{Gell-Mann:1979vob}%
  \BibitemOpen
  \bibfield  {author} {\bibinfo {author} {\bibfnamefont {M.}~\bibnamefont
  {Gell-Mann}}, \bibinfo {author} {\bibfnamefont {P.}~\bibnamefont {Ramond}}, \
  and\ \bibinfo {author} {\bibfnamefont {R.}~\bibnamefont {Slansky}},\
  }\href@noop {} {\bibfield  {journal} {\bibinfo  {journal} {Conf. Proc. C}\
  }\textbf {\bibinfo {volume} {790927}},\ \bibinfo {pages} {315} (\bibinfo
  {year} {1979})},\ \Eprint {http://arxiv.org/abs/1306.4669} {arXiv:1306.4669
  [hep-th]} \BibitemShut {NoStop}%
\bibitem [{\citenamefont {Fukugita}\ and\ \citenamefont
  {Yanagida}(1986)}]{Fukugita:1986hr}%
  \BibitemOpen
  \bibfield  {author} {\bibinfo {author} {\bibfnamefont {M.}~\bibnamefont
  {Fukugita}}\ and\ \bibinfo {author} {\bibfnamefont {T.}~\bibnamefont
  {Yanagida}},\ }\href {\doibase 10.1016/0370-2693(86)91126-3} {\bibfield
  {journal} {\bibinfo  {journal} {Phys. Lett. B}\ }\textbf {\bibinfo {volume}
  {174}},\ \bibinfo {pages} {45} (\bibinfo {year} {1986})}\BibitemShut
  {NoStop}%
\bibitem [{\citenamefont {Magg}\ and\ \citenamefont
  {Wetterich}(1980)}]{Magg:1980ut}%
  \BibitemOpen
  \bibfield  {author} {\bibinfo {author} {\bibfnamefont {M.}~\bibnamefont
  {Magg}}\ and\ \bibinfo {author} {\bibfnamefont {C.}~\bibnamefont
  {Wetterich}},\ }\href {\doibase 10.1016/0370-2693(80)90825-4} {\bibfield
  {journal} {\bibinfo  {journal} {Phys. Lett. B}\ }\textbf {\bibinfo {volume}
  {94}},\ \bibinfo {pages} {61} (\bibinfo {year} {1980})}\BibitemShut {NoStop}%
\bibitem [{\citenamefont {'t~Hooft}(1974)}]{tHooft:1974kcl}%
  \BibitemOpen
  \bibfield  {author} {\bibinfo {author} {\bibfnamefont {G.}~\bibnamefont
  {'t~Hooft}},\ }\href {\doibase 10.1016/0550-3213(74)90486-6} {\bibfield
  {journal} {\bibinfo  {journal} {Nucl. Phys. B}\ }\textbf {\bibinfo {volume}
  {79}},\ \bibinfo {pages} {276} (\bibinfo {year} {1974})}\BibitemShut
  {NoStop}%
\bibitem [{\citenamefont {Polyakov}(1974)}]{Polyakov:1974ek}%
  \BibitemOpen
  \bibfield  {author} {\bibinfo {author} {\bibfnamefont {A.~M.}\ \bibnamefont
  {Polyakov}},\ }\href@noop {} {\bibfield  {journal} {\bibinfo  {journal} {JETP
  Lett.}\ }\textbf {\bibinfo {volume} {20}},\ \bibinfo {pages} {194} (\bibinfo
  {year} {1974})}\BibitemShut {NoStop}%
\bibitem [{\citenamefont {Kibble}(1976)}]{Kibble:1976sj}%
  \BibitemOpen
  \bibfield  {author} {\bibinfo {author} {\bibfnamefont {T.~W.~B.}\
  \bibnamefont {Kibble}},\ }\href {\doibase 10.1088/0305-4470/9/8/029}
  {\bibfield  {journal} {\bibinfo  {journal} {J. Phys. A}\ }\textbf {\bibinfo
  {volume} {9}},\ \bibinfo {pages} {1387} (\bibinfo {year} {1976})}\BibitemShut
  {NoStop}%
\bibitem [{\citenamefont {Turok}(1989)}]{Turok:1989ai}%
  \BibitemOpen
  \bibfield  {author} {\bibinfo {author} {\bibfnamefont {N.}~\bibnamefont
  {Turok}},\ }\href {\doibase 10.1103/PhysRevLett.63.2625} {\bibfield
  {journal} {\bibinfo  {journal} {Phys. Rev. Lett.}\ }\textbf {\bibinfo
  {volume} {63}},\ \bibinfo {pages} {2625} (\bibinfo {year}
  {1989})}\BibitemShut {NoStop}%
\bibitem [{\citenamefont {Okada}\ and\ \citenamefont
  {Shafi}(2017)}]{Okada:2015vka}%
  \BibitemOpen
  \bibfield  {author} {\bibinfo {author} {\bibfnamefont {N.}~\bibnamefont
  {Okada}}\ and\ \bibinfo {author} {\bibfnamefont {Q.}~\bibnamefont {Shafi}},\
  }\href {\doibase 10.1016/j.physletb.2017.11.015} {\bibfield  {journal}
  {\bibinfo  {journal} {Phys. Lett. B}\ }\textbf {\bibinfo {volume} {775}},\
  \bibinfo {pages} {348} (\bibinfo {year} {2017})},\ \Eprint
  {http://arxiv.org/abs/1506.01410} {arXiv:1506.01410 [hep-ph]} \BibitemShut
  {NoStop}%
\bibitem [{\citenamefont {Dvali}\ \emph {et~al.}(1998)\citenamefont {Dvali},
  \citenamefont {Lazarides},\ and\ \citenamefont {Shafi}}]{Dvali:1997uq}%
  \BibitemOpen
  \bibfield  {author} {\bibinfo {author} {\bibfnamefont {G.~R.}\ \bibnamefont
  {Dvali}}, \bibinfo {author} {\bibfnamefont {G.}~\bibnamefont {Lazarides}}, \
  and\ \bibinfo {author} {\bibfnamefont {Q.}~\bibnamefont {Shafi}},\ }\href
  {\doibase 10.1016/S0370-2693(98)00145-2} {\bibfield  {journal} {\bibinfo
  {journal} {Phys. Lett. B}\ }\textbf {\bibinfo {volume} {424}},\ \bibinfo
  {pages} {259} (\bibinfo {year} {1998})},\ \Eprint
  {http://arxiv.org/abs/hep-ph/9710314} {arXiv:hep-ph/9710314} \BibitemShut
  {NoStop}%
\bibitem [{\citenamefont {Lazarides}\ \emph
  {et~al.}(2023{\natexlab{a}})\citenamefont {Lazarides}, \citenamefont
  {Shafi},\ and\ \citenamefont {Tiwari}}]{Lazarides:2023iim}%
  \BibitemOpen
  \bibfield  {author} {\bibinfo {author} {\bibfnamefont {G.}~\bibnamefont
  {Lazarides}}, \bibinfo {author} {\bibfnamefont {Q.}~\bibnamefont {Shafi}}, \
  and\ \bibinfo {author} {\bibfnamefont {A.}~\bibnamefont {Tiwari}},\ }\href
  {\doibase 10.1007/JHEP05(2023)119} {\bibfield  {journal} {\bibinfo  {journal}
  {JHEP}\ }\textbf {\bibinfo {volume} {05}},\ \bibinfo {pages} {119} (\bibinfo
  {year} {2023}{\natexlab{a}})},\ \Eprint {http://arxiv.org/abs/2303.15159}
  {arXiv:2303.15159 [hep-ph]} \BibitemShut {NoStop}%
\bibitem [{\citenamefont {Lazarides}\ \emph {et~al.}(2020)\citenamefont
  {Lazarides}, \citenamefont {Rehman},\ and\ \citenamefont
  {Shafi}}]{Lazarides:2020bgy}%
  \BibitemOpen
  \bibfield  {author} {\bibinfo {author} {\bibfnamefont {G.}~\bibnamefont
  {Lazarides}}, \bibinfo {author} {\bibfnamefont {M.~U.}\ \bibnamefont
  {Rehman}}, \ and\ \bibinfo {author} {\bibfnamefont {Q.}~\bibnamefont
  {Shafi}},\ }\href {\doibase 10.1007/JHEP10(2020)085} {\bibfield  {journal}
  {\bibinfo  {journal} {JHEP}\ }\textbf {\bibinfo {volume} {10}},\ \bibinfo
  {pages} {085} (\bibinfo {year} {2020})},\ \Eprint
  {http://arxiv.org/abs/2007.15317} {arXiv:2007.15317 [hep-ph]} \BibitemShut
  {NoStop}%
\bibitem [{\citenamefont {Mehmood}\ \emph {et~al.}(2021)\citenamefont
  {Mehmood}, \citenamefont {Rehman},\ and\ \citenamefont
  {Shafi}}]{Mehmood:2020irm}%
  \BibitemOpen
  \bibfield  {author} {\bibinfo {author} {\bibfnamefont {M.}~\bibnamefont
  {Mehmood}}, \bibinfo {author} {\bibfnamefont {M.~U.}\ \bibnamefont {Rehman}},
  \ and\ \bibinfo {author} {\bibfnamefont {Q.}~\bibnamefont {Shafi}},\ }\href
  {\doibase 10.1007/JHEP02(2021)181} {\bibfield  {journal} {\bibinfo  {journal}
  {JHEP}\ }\textbf {\bibinfo {volume} {02}},\ \bibinfo {pages} {181} (\bibinfo
  {year} {2021})},\ \Eprint {http://arxiv.org/abs/2010.01665} {arXiv:2010.01665
  [hep-ph]} \BibitemShut {NoStop}%
\bibitem [{\citenamefont {Abid}\ \emph {et~al.}(2021)\citenamefont {Abid},
  \citenamefont {Mehmood}, \citenamefont {Rehman},\ and\ \citenamefont
  {Shafi}}]{Abid:2021jvn}%
  \BibitemOpen
  \bibfield  {author} {\bibinfo {author} {\bibfnamefont {M.~M.~A.}\
  \bibnamefont {Abid}}, \bibinfo {author} {\bibfnamefont {M.}~\bibnamefont
  {Mehmood}}, \bibinfo {author} {\bibfnamefont {M.~U.}\ \bibnamefont {Rehman}},
  \ and\ \bibinfo {author} {\bibfnamefont {Q.}~\bibnamefont {Shafi}},\ }\href
  {\doibase 10.1088/1475-7516/2021/10/015} {\bibfield  {journal} {\bibinfo
  {journal} {JCAP}\ }\textbf {\bibinfo {volume} {10}},\ \bibinfo {pages} {015}
  (\bibinfo {year} {2021})},\ \Eprint {http://arxiv.org/abs/2107.05678}
  {arXiv:2107.05678 [hep-ph]} \BibitemShut {NoStop}%
\bibitem [{\citenamefont {Mehmood}\ and\ \citenamefont
  {Rehman}(2023)}]{Mehmood:2023gmm}%
  \BibitemOpen
  \bibfield  {author} {\bibinfo {author} {\bibfnamefont {M.}~\bibnamefont
  {Mehmood}}\ and\ \bibinfo {author} {\bibfnamefont {M.~U.}\ \bibnamefont
  {Rehman}},\ }\href {\doibase 10.1103/PhysRevD.108.075030} {\bibfield
  {journal} {\bibinfo  {journal} {Phys. Rev. D}\ }\textbf {\bibinfo {volume}
  {108}},\ \bibinfo {pages} {075030} (\bibinfo {year} {2023})},\ \Eprint
  {http://arxiv.org/abs/2305.00611} {arXiv:2305.00611 [hep-ph]} \BibitemShut
  {NoStop}%
\bibitem [{\citenamefont {Ijaz}\ \emph {et~al.}(2023)\citenamefont {Ijaz},
  \citenamefont {Mehmood},\ and\ \citenamefont {Rehman}}]{Ijaz:2023cvc}%
  \BibitemOpen
  \bibfield  {author} {\bibinfo {author} {\bibfnamefont {N.}~\bibnamefont
  {Ijaz}}, \bibinfo {author} {\bibfnamefont {M.}~\bibnamefont {Mehmood}}, \
  and\ \bibinfo {author} {\bibfnamefont {M.~U.}\ \bibnamefont {Rehman}},\
  }\href@noop {} {\  (\bibinfo {year} {2023})},\ \Eprint
  {http://arxiv.org/abs/2308.14908} {arXiv:2308.14908 [astro-ph.CO]}
  \BibitemShut {NoStop}%
\bibitem [{\citenamefont {Abe}\ \emph {et~al.}(2018)\citenamefont {Abe} \emph
  {et~al.}}]{Hyper-Kamiokande:2018ofw}%
  \BibitemOpen
  \bibfield  {author} {\bibinfo {author} {\bibfnamefont {K.}~\bibnamefont
  {Abe}} \emph {et~al.} (\bibinfo {collaboration} {Hyper-Kamiokande}),\
  }\href@noop {} {\  (\bibinfo {year} {2018})},\ \Eprint
  {http://arxiv.org/abs/1805.04163} {arXiv:1805.04163 [physics.ins-det]}
  \BibitemShut {NoStop}%
\bibitem [{\citenamefont {Abi}\ \emph {et~al.}(2020)\citenamefont {Abi} \emph
  {et~al.}}]{DUNE:2020ypp}%
  \BibitemOpen
  \bibfield  {author} {\bibinfo {author} {\bibfnamefont {B.}~\bibnamefont
  {Abi}} \emph {et~al.} (\bibinfo {collaboration} {DUNE}),\ }\href@noop {} {\
  (\bibinfo {year} {2020})},\ \Eprint {http://arxiv.org/abs/2002.03005}
  {arXiv:2002.03005 [hep-ex]} \BibitemShut {NoStop}%
\bibitem [{\citenamefont {Acciarri}\ \emph {et~al.}(2015)\citenamefont
  {Acciarri} \emph {et~al.}}]{DUNE:2015lol}%
  \BibitemOpen
  \bibfield  {author} {\bibinfo {author} {\bibfnamefont {R.}~\bibnamefont
  {Acciarri}} \emph {et~al.} (\bibinfo {collaboration} {DUNE}),\ }\href@noop {}
  {\  (\bibinfo {year} {2015})},\ \Eprint {http://arxiv.org/abs/1512.06148}
  {arXiv:1512.06148 [physics.ins-det]} \BibitemShut {NoStop}%
\bibitem [{\citenamefont {Rehman}\ \emph {et~al.}(2017)\citenamefont {Rehman},
  \citenamefont {Shafi},\ and\ \citenamefont {Vardag}}]{Rehman:2017gkm}%
  \BibitemOpen
  \bibfield  {author} {\bibinfo {author} {\bibfnamefont {M.~U.}\ \bibnamefont
  {Rehman}}, \bibinfo {author} {\bibfnamefont {Q.}~\bibnamefont {Shafi}}, \
  and\ \bibinfo {author} {\bibfnamefont {F.~K.}\ \bibnamefont {Vardag}},\
  }\href {\doibase 10.1103/PhysRevD.96.063527} {\bibfield  {journal} {\bibinfo
  {journal} {Phys. Rev. D}\ }\textbf {\bibinfo {volume} {96}},\ \bibinfo
  {pages} {063527} (\bibinfo {year} {2017})},\ \Eprint
  {http://arxiv.org/abs/1705.03693} {arXiv:1705.03693 [hep-ph]} \BibitemShut
  {NoStop}%
\bibitem [{\citenamefont {Okada}\ and\ \citenamefont
  {Shafi}(2018)}]{Okada:2017rbf}%
  \BibitemOpen
  \bibfield  {author} {\bibinfo {author} {\bibfnamefont {N.}~\bibnamefont
  {Okada}}\ and\ \bibinfo {author} {\bibfnamefont {Q.}~\bibnamefont {Shafi}},\
  }\href {\doibase 10.1016/j.physletb.2018.10.057} {\bibfield  {journal}
  {\bibinfo  {journal} {Phys. Lett. B}\ }\textbf {\bibinfo {volume} {787}},\
  \bibinfo {pages} {141} (\bibinfo {year} {2018})},\ \Eprint
  {http://arxiv.org/abs/1709.04610} {arXiv:1709.04610 [hep-ph]} \BibitemShut
  {NoStop}%
\bibitem [{\citenamefont {Lazarides}\ \emph {et~al.}(2021)\citenamefont
  {Lazarides}, \citenamefont {Rehman}, \citenamefont {Shafi},\ and\
  \citenamefont {Vardag}}]{Lazarides:2020zof}%
  \BibitemOpen
  \bibfield  {author} {\bibinfo {author} {\bibfnamefont {G.}~\bibnamefont
  {Lazarides}}, \bibinfo {author} {\bibfnamefont {M.~U.}\ \bibnamefont
  {Rehman}}, \bibinfo {author} {\bibfnamefont {Q.}~\bibnamefont {Shafi}}, \
  and\ \bibinfo {author} {\bibfnamefont {F.~K.}\ \bibnamefont {Vardag}},\
  }\href {\doibase 10.1103/PhysRevD.103.035033} {\bibfield  {journal} {\bibinfo
   {journal} {Phys. Rev. D}\ }\textbf {\bibinfo {volume} {103}},\ \bibinfo
  {pages} {035033} (\bibinfo {year} {2021})},\ \Eprint
  {http://arxiv.org/abs/2007.01474} {arXiv:2007.01474 [hep-ph]} \BibitemShut
  {NoStop}%
\bibitem [{\citenamefont {Afzal}\ \emph {et~al.}(2022)\citenamefont {Afzal},
  \citenamefont {Ahmed}, \citenamefont {Rehman},\ and\ \citenamefont
  {Shafi}}]{Afzal:2022vjx}%
  \BibitemOpen
  \bibfield  {author} {\bibinfo {author} {\bibfnamefont {A.}~\bibnamefont
  {Afzal}}, \bibinfo {author} {\bibfnamefont {W.}~\bibnamefont {Ahmed}},
  \bibinfo {author} {\bibfnamefont {M.~U.}\ \bibnamefont {Rehman}}, \ and\
  \bibinfo {author} {\bibfnamefont {Q.}~\bibnamefont {Shafi}},\ }\href
  {\doibase 10.1103/PhysRevD.105.103539} {\bibfield  {journal} {\bibinfo
  {journal} {Phys. Rev. D}\ }\textbf {\bibinfo {volume} {105}},\ \bibinfo
  {pages} {103539} (\bibinfo {year} {2022})},\ \Eprint
  {http://arxiv.org/abs/2202.07386} {arXiv:2202.07386 [hep-ph]} \BibitemShut
  {NoStop}%
\bibitem [{\citenamefont {Afzal}\ \emph
  {et~al.}(2023{\natexlab{b}})\citenamefont {Afzal}, \citenamefont {Mehmood},
  \citenamefont {Rehman},\ and\ \citenamefont {Shafi}}]{Afzal:2023cyp}%
  \BibitemOpen
  \bibfield  {author} {\bibinfo {author} {\bibfnamefont {A.}~\bibnamefont
  {Afzal}}, \bibinfo {author} {\bibfnamefont {M.}~\bibnamefont {Mehmood}},
  \bibinfo {author} {\bibfnamefont {M.~U.}\ \bibnamefont {Rehman}}, \ and\
  \bibinfo {author} {\bibfnamefont {Q.}~\bibnamefont {Shafi}},\ }\href@noop {}
  {\  (\bibinfo {year} {2023}{\natexlab{b}})},\ \Eprint
  {http://arxiv.org/abs/2308.11410} {arXiv:2308.11410 [hep-ph]} \BibitemShut
  {NoStop}%
\bibitem [{\citenamefont {Zubair}(2024)}]{Zubair:2024quc}%
  \BibitemOpen
  \bibfield  {author} {\bibinfo {author} {\bibfnamefont {U.}~\bibnamefont
  {Zubair}},\ }\href@noop {} {\  (\bibinfo {year} {2024})},\ \Eprint
  {http://arxiv.org/abs/2403.13991} {arXiv:2403.13991 [hep-ph]} \BibitemShut
  {NoStop}%
\bibitem [{\citenamefont {Dine}\ \emph
  {et~al.}(1987{\natexlab{a}})\citenamefont {Dine}, \citenamefont {Seiberg},\
  and\ \citenamefont {Witten}}]{Dine:1987xk}%
  \BibitemOpen
  \bibfield  {author} {\bibinfo {author} {\bibfnamefont {M.}~\bibnamefont
  {Dine}}, \bibinfo {author} {\bibfnamefont {N.}~\bibnamefont {Seiberg}}, \
  and\ \bibinfo {author} {\bibfnamefont {E.}~\bibnamefont {Witten}},\ }\href
  {\doibase 10.1016/0550-3213(87)90395-6} {\bibfield  {journal} {\bibinfo
  {journal} {Nucl. Phys. B}\ }\textbf {\bibinfo {volume} {289}},\ \bibinfo
  {pages} {589} (\bibinfo {year} {1987}{\natexlab{a}})}\BibitemShut {NoStop}%
\bibitem [{\citenamefont {Atick}\ \emph {et~al.}(1987)\citenamefont {Atick},
  \citenamefont {Dixon},\ and\ \citenamefont {Sen}}]{Atick:1987gy}%
  \BibitemOpen
  \bibfield  {author} {\bibinfo {author} {\bibfnamefont {J.~J.}\ \bibnamefont
  {Atick}}, \bibinfo {author} {\bibfnamefont {L.~J.}\ \bibnamefont {Dixon}}, \
  and\ \bibinfo {author} {\bibfnamefont {A.}~\bibnamefont {Sen}},\ }\href
  {\doibase 10.1016/0550-3213(87)90639-0} {\bibfield  {journal} {\bibinfo
  {journal} {Nucl. Phys. B}\ }\textbf {\bibinfo {volume} {292}},\ \bibinfo
  {pages} {109} (\bibinfo {year} {1987})}\BibitemShut {NoStop}%
\bibitem [{\citenamefont {Dine}\ \emph
  {et~al.}(1987{\natexlab{b}})\citenamefont {Dine}, \citenamefont {Ichinose},\
  and\ \citenamefont {Seiberg}}]{Dine:1987gj}%
  \BibitemOpen
  \bibfield  {author} {\bibinfo {author} {\bibfnamefont {M.}~\bibnamefont
  {Dine}}, \bibinfo {author} {\bibfnamefont {I.}~\bibnamefont {Ichinose}}, \
  and\ \bibinfo {author} {\bibfnamefont {N.}~\bibnamefont {Seiberg}},\ }\href
  {\doibase 10.1016/0550-3213(87)90072-1} {\bibfield  {journal} {\bibinfo
  {journal} {Nucl. Phys. B}\ }\textbf {\bibinfo {volume} {293}},\ \bibinfo
  {pages} {253} (\bibinfo {year} {1987}{\natexlab{b}})}\BibitemShut {NoStop}%
\bibitem [{\citenamefont {Copeland}\ \emph {et~al.}(1994)\citenamefont
  {Copeland}, \citenamefont {Liddle}, \citenamefont {Lyth}, \citenamefont
  {Stewart},\ and\ \citenamefont {Wands}}]{Copeland:1994vg}%
  \BibitemOpen
  \bibfield  {author} {\bibinfo {author} {\bibfnamefont {E.~J.}\ \bibnamefont
  {Copeland}}, \bibinfo {author} {\bibfnamefont {A.~R.}\ \bibnamefont
  {Liddle}}, \bibinfo {author} {\bibfnamefont {D.~H.}\ \bibnamefont {Lyth}},
  \bibinfo {author} {\bibfnamefont {E.~D.}\ \bibnamefont {Stewart}}, \ and\
  \bibinfo {author} {\bibfnamefont {D.}~\bibnamefont {Wands}},\ }\href
  {\doibase 10.1103/PhysRevD.49.6410} {\bibfield  {journal} {\bibinfo
  {journal} {Phys. Rev. D}\ }\textbf {\bibinfo {volume} {49}},\ \bibinfo
  {pages} {6410} (\bibinfo {year} {1994})},\ \Eprint
  {http://arxiv.org/abs/astro-ph/9401011} {arXiv:astro-ph/9401011} \BibitemShut
  {NoStop}%
\bibitem [{\citenamefont {Dvali}\ \emph {et~al.}(1994)\citenamefont {Dvali},
  \citenamefont {Shafi},\ and\ \citenamefont {Schaefer}}]{Dvali:1994ms}%
  \BibitemOpen
  \bibfield  {author} {\bibinfo {author} {\bibfnamefont {G.~R.}\ \bibnamefont
  {Dvali}}, \bibinfo {author} {\bibfnamefont {Q.}~\bibnamefont {Shafi}}, \ and\
  \bibinfo {author} {\bibfnamefont {R.~K.}\ \bibnamefont {Schaefer}},\ }\href
  {\doibase 10.1103/PhysRevLett.73.1886} {\bibfield  {journal} {\bibinfo
  {journal} {Phys. Rev. Lett.}\ }\textbf {\bibinfo {volume} {73}},\ \bibinfo
  {pages} {1886} (\bibinfo {year} {1994})},\ \Eprint
  {http://arxiv.org/abs/hep-ph/9406319} {arXiv:hep-ph/9406319} \BibitemShut
  {NoStop}%
\bibitem [{\citenamefont {Senoguz}\ and\ \citenamefont
  {Shafi}(2005)}]{Senoguz:2004vu}%
  \BibitemOpen
  \bibfield  {author} {\bibinfo {author} {\bibfnamefont {V.~N.}\ \bibnamefont
  {Senoguz}}\ and\ \bibinfo {author} {\bibfnamefont {Q.}~\bibnamefont
  {Shafi}},\ }\href {\doibase 10.1103/PhysRevD.71.043514} {\bibfield  {journal}
  {\bibinfo  {journal} {Phys. Rev. D}\ }\textbf {\bibinfo {volume} {71}},\
  \bibinfo {pages} {043514} (\bibinfo {year} {2005})},\ \Eprint
  {http://arxiv.org/abs/hep-ph/0412102} {arXiv:hep-ph/0412102} \BibitemShut
  {NoStop}%
\bibitem [{\citenamefont {Buchmuller}\ \emph {et~al.}(2000)\citenamefont
  {Buchmuller}, \citenamefont {Covi},\ and\ \citenamefont
  {Delepine}}]{Buchmuller:2000zm}%
  \BibitemOpen
  \bibfield  {author} {\bibinfo {author} {\bibfnamefont {W.}~\bibnamefont
  {Buchmuller}}, \bibinfo {author} {\bibfnamefont {L.}~\bibnamefont {Covi}}, \
  and\ \bibinfo {author} {\bibfnamefont {D.}~\bibnamefont {Delepine}},\ }\href
  {\doibase 10.1016/S0370-2693(00)01005-4} {\bibfield  {journal} {\bibinfo
  {journal} {Phys. Lett. B}\ }\textbf {\bibinfo {volume} {491}},\ \bibinfo
  {pages} {183} (\bibinfo {year} {2000})},\ \Eprint
  {http://arxiv.org/abs/hep-ph/0006168} {arXiv:hep-ph/0006168} \BibitemShut
  {NoStop}%
\bibitem [{\citenamefont {Rehman}\ \emph {et~al.}(2010)\citenamefont {Rehman},
  \citenamefont {Shafi},\ and\ \citenamefont {Wickman}}]{Rehman:2009nq}%
  \BibitemOpen
  \bibfield  {author} {\bibinfo {author} {\bibfnamefont {M.~U.}\ \bibnamefont
  {Rehman}}, \bibinfo {author} {\bibfnamefont {Q.}~\bibnamefont {Shafi}}, \
  and\ \bibinfo {author} {\bibfnamefont {J.~R.}\ \bibnamefont {Wickman}},\
  }\href {\doibase 10.1016/j.physletb.2009.12.010} {\bibfield  {journal}
  {\bibinfo  {journal} {Phys. Lett. B}\ }\textbf {\bibinfo {volume} {683}},\
  \bibinfo {pages} {191} (\bibinfo {year} {2010})},\ \Eprint
  {http://arxiv.org/abs/0908.3896} {arXiv:0908.3896 [hep-ph]} \BibitemShut
  {NoStop}%
\bibitem [{\citenamefont {ur~Rehman}\ \emph {et~al.}(2007)\citenamefont
  {ur~Rehman}, \citenamefont {Senoguz},\ and\ \citenamefont
  {Shafi}}]{urRehman:2006hu}%
  \BibitemOpen
  \bibfield  {author} {\bibinfo {author} {\bibfnamefont {M.}~\bibnamefont
  {ur~Rehman}}, \bibinfo {author} {\bibfnamefont {V.~N.}\ \bibnamefont
  {Senoguz}}, \ and\ \bibinfo {author} {\bibfnamefont {Q.}~\bibnamefont
  {Shafi}},\ }\href {\doibase 10.1103/PhysRevD.75.043522} {\bibfield  {journal}
  {\bibinfo  {journal} {Phys. Rev. D}\ }\textbf {\bibinfo {volume} {75}},\
  \bibinfo {pages} {043522} (\bibinfo {year} {2007})},\ \Eprint
  {http://arxiv.org/abs/hep-ph/0612023} {arXiv:hep-ph/0612023} \BibitemShut
  {NoStop}%
\bibitem [{\citenamefont {Rehman}\ \emph {et~al.}(2011)\citenamefont {Rehman},
  \citenamefont {Shafi},\ and\ \citenamefont {Wickman}}]{Rehman:2010wm}%
  \BibitemOpen
  \bibfield  {author} {\bibinfo {author} {\bibfnamefont {M.~U.}\ \bibnamefont
  {Rehman}}, \bibinfo {author} {\bibfnamefont {Q.}~\bibnamefont {Shafi}}, \
  and\ \bibinfo {author} {\bibfnamefont {J.~R.}\ \bibnamefont {Wickman}},\
  }\href {\doibase 10.1103/PhysRevD.83.067304} {\bibfield  {journal} {\bibinfo
  {journal} {Phys. Rev. D}\ }\textbf {\bibinfo {volume} {83}},\ \bibinfo
  {pages} {067304} (\bibinfo {year} {2011})},\ \Eprint
  {http://arxiv.org/abs/1012.0309} {arXiv:1012.0309 [astro-ph.CO]} \BibitemShut
  {NoStop}%
\bibitem [{\citenamefont {Buchm\"uller}\ \emph {et~al.}(2014)\citenamefont
  {Buchm\"uller}, \citenamefont {Domcke}, \citenamefont {Kamada},\ and\
  \citenamefont {Schmitz}}]{Buchmuller:2014epa}%
  \BibitemOpen
  \bibfield  {author} {\bibinfo {author} {\bibfnamefont {W.}~\bibnamefont
  {Buchm\"uller}}, \bibinfo {author} {\bibfnamefont {V.}~\bibnamefont
  {Domcke}}, \bibinfo {author} {\bibfnamefont {K.}~\bibnamefont {Kamada}}, \
  and\ \bibinfo {author} {\bibfnamefont {K.}~\bibnamefont {Schmitz}},\ }\href
  {\doibase 10.1088/1475-7516/2014/07/054} {\bibfield  {journal} {\bibinfo
  {journal} {JCAP}\ }\textbf {\bibinfo {volume} {07}},\ \bibinfo {pages} {054}
  (\bibinfo {year} {2014})},\ \Eprint {http://arxiv.org/abs/1404.1832}
  {arXiv:1404.1832 [hep-ph]} \BibitemShut {NoStop}%
\bibitem [{\citenamefont {Lazarides}\ and\ \citenamefont
  {Vlachos}(1998)}]{Lazarides:1998qx}%
  \BibitemOpen
  \bibfield  {author} {\bibinfo {author} {\bibfnamefont {G.}~\bibnamefont
  {Lazarides}}\ and\ \bibinfo {author} {\bibfnamefont {N.~D.}\ \bibnamefont
  {Vlachos}},\ }\href {\doibase 10.1016/S0370-2693(98)01126-5} {\bibfield
  {journal} {\bibinfo  {journal} {Phys. Lett. B}\ }\textbf {\bibinfo {volume}
  {441}},\ \bibinfo {pages} {46} (\bibinfo {year} {1998})},\ \Eprint
  {http://arxiv.org/abs/hep-ph/9807253} {arXiv:hep-ph/9807253} \BibitemShut
  {NoStop}%
\bibitem [{\citenamefont {Liddle}\ and\ \citenamefont
  {Leach}(2003)}]{Liddle:2003as}%
  \BibitemOpen
  \bibfield  {author} {\bibinfo {author} {\bibfnamefont {A.~R.}\ \bibnamefont
  {Liddle}}\ and\ \bibinfo {author} {\bibfnamefont {S.~M.}\ \bibnamefont
  {Leach}},\ }\href {\doibase 10.1103/PhysRevD.68.103503} {\bibfield  {journal}
  {\bibinfo  {journal} {Phys. Rev. D}\ }\textbf {\bibinfo {volume} {68}},\
  \bibinfo {pages} {103503} (\bibinfo {year} {2003})},\ \Eprint
  {http://arxiv.org/abs/astro-ph/0305263} {arXiv:astro-ph/0305263} \BibitemShut
  {NoStop}%
\bibitem [{\citenamefont {Kuzmin}\ \emph {et~al.}(1985)\citenamefont {Kuzmin},
  \citenamefont {Rubakov},\ and\ \citenamefont {Shaposhnikov}}]{Kuzmin:1985mm}%
  \BibitemOpen
  \bibfield  {author} {\bibinfo {author} {\bibfnamefont {V.~A.}\ \bibnamefont
  {Kuzmin}}, \bibinfo {author} {\bibfnamefont {V.~A.}\ \bibnamefont {Rubakov}},
  \ and\ \bibinfo {author} {\bibfnamefont {M.~E.}\ \bibnamefont
  {Shaposhnikov}},\ }\href {\doibase 10.1016/0370-2693(85)91028-7} {\bibfield
  {journal} {\bibinfo  {journal} {Phys. Lett. B}\ }\textbf {\bibinfo {volume}
  {155}},\ \bibinfo {pages} {36} (\bibinfo {year} {1985})}\BibitemShut
  {NoStop}%
\bibitem [{\citenamefont {Khlebnikov}\ and\ \citenamefont
  {Shaposhnikov}(1988)}]{Khlebnikov:1988sr}%
  \BibitemOpen
  \bibfield  {author} {\bibinfo {author} {\bibfnamefont {S.~Y.}\ \bibnamefont
  {Khlebnikov}}\ and\ \bibinfo {author} {\bibfnamefont {M.~E.}\ \bibnamefont
  {Shaposhnikov}},\ }\href {\doibase 10.1016/0550-3213(88)90133-2} {\bibfield
  {journal} {\bibinfo  {journal} {Nucl. Phys. B}\ }\textbf {\bibinfo {volume}
  {308}},\ \bibinfo {pages} {885} (\bibinfo {year} {1988})}\BibitemShut
  {NoStop}%
\bibitem [{\citenamefont {Rehman}\ \emph {et~al.}(2020)\citenamefont {Rehman},
  \citenamefont {Abid},\ and\ \citenamefont {Ejaz}}]{Rehman:2018gnr}%
  \BibitemOpen
  \bibfield  {author} {\bibinfo {author} {\bibfnamefont {M.~U.}\ \bibnamefont
  {Rehman}}, \bibinfo {author} {\bibfnamefont {M.~M.~A.}\ \bibnamefont {Abid}},
  \ and\ \bibinfo {author} {\bibfnamefont {A.}~\bibnamefont {Ejaz}},\ }\href
  {\doibase 10.1088/1475-7516/2020/11/019} {\bibfield  {journal} {\bibinfo
  {journal} {JCAP}\ }\textbf {\bibinfo {volume} {11}},\ \bibinfo {pages} {019}
  (\bibinfo {year} {2020})},\ \Eprint {http://arxiv.org/abs/1804.07619}
  {arXiv:1804.07619 [hep-ph]} \BibitemShut {NoStop}%
\bibitem [{\citenamefont {Zyla}\ \emph {et~al.}(2020)\citenamefont {Zyla} \emph
  {et~al.}}]{ParticleDataGroup:2020ssz}%
  \BibitemOpen
  \bibfield  {author} {\bibinfo {author} {\bibfnamefont {P.~A.}\ \bibnamefont
  {Zyla}} \emph {et~al.} (\bibinfo {collaboration} {Particle Data Group}),\
  }\href {\doibase 10.1093/ptep/ptaa104} {\bibfield  {journal} {\bibinfo
  {journal} {PTEP}\ }\textbf {\bibinfo {volume} {2020}},\ \bibinfo {pages}
  {083C01} (\bibinfo {year} {2020})}\BibitemShut {NoStop}%
\bibitem [{\citenamefont {Akrami}\ \emph {et~al.}(2020)\citenamefont {Akrami}
  \emph {et~al.}}]{Planck:2018jri}%
  \BibitemOpen
  \bibfield  {author} {\bibinfo {author} {\bibfnamefont {Y.}~\bibnamefont
  {Akrami}} \emph {et~al.} (\bibinfo {collaboration} {Planck}),\ }\href
  {\doibase 10.1051/0004-6361/201833887} {\bibfield  {journal} {\bibinfo
  {journal} {Astron. Astrophys.}\ }\textbf {\bibinfo {volume} {641}},\ \bibinfo
  {pages} {A10} (\bibinfo {year} {2020})},\ \Eprint
  {http://arxiv.org/abs/1807.06211} {arXiv:1807.06211 [astro-ph.CO]}
  \BibitemShut {NoStop}%
\bibitem [{\citenamefont {Ferdman}\ \emph {et~al.}(2010)\citenamefont {Ferdman}
  \emph {et~al.}}]{Ferdman:2010xq}%
  \BibitemOpen
  \bibfield  {author} {\bibinfo {author} {\bibfnamefont {R.~D.}\ \bibnamefont
  {Ferdman}} \emph {et~al.},\ }\href {\doibase 10.1088/0264-9381/27/8/084014}
  {\bibfield  {journal} {\bibinfo  {journal} {Class. Quant. Grav.}\ }\textbf
  {\bibinfo {volume} {27}},\ \bibinfo {pages} {084014} (\bibinfo {year}
  {2010})},\ \Eprint {http://arxiv.org/abs/1003.3405} {arXiv:1003.3405
  [astro-ph.HE]} \BibitemShut {NoStop}%
\bibitem [{\citenamefont {Amaro-Seoane}\ \emph {et~al.}(2017)\citenamefont
  {Amaro-Seoane} \emph {et~al.}}]{LISA:2017pwj}%
  \BibitemOpen
  \bibfield  {author} {\bibinfo {author} {\bibfnamefont {P.}~\bibnamefont
  {Amaro-Seoane}} \emph {et~al.} (\bibinfo {collaboration} {LISA}),\
  }\href@noop {} {\  (\bibinfo {year} {2017})},\ \Eprint
  {http://arxiv.org/abs/1702.00786} {arXiv:1702.00786 [astro-ph.IM]}
  \BibitemShut {NoStop}%
\bibitem [{\citenamefont {Hu}\ and\ \citenamefont {Wu}(2017)}]{Hu:2017mde}%
  \BibitemOpen
  \bibfield  {author} {\bibinfo {author} {\bibfnamefont {W.-R.}\ \bibnamefont
  {Hu}}\ and\ \bibinfo {author} {\bibfnamefont {Y.-L.}\ \bibnamefont {Wu}},\
  }\href {\doibase 10.1093/nsr/nwx116} {\bibfield  {journal} {\bibinfo
  {journal} {Natl. Sci. Rev.}\ }\textbf {\bibinfo {volume} {4}},\ \bibinfo
  {pages} {685} (\bibinfo {year} {2017})}\BibitemShut {NoStop}%
\bibitem [{\citenamefont {Luo}\ \emph {et~al.}(2016)\citenamefont {Luo} \emph
  {et~al.}}]{TianQin:2015yph}%
  \BibitemOpen
  \bibfield  {author} {\bibinfo {author} {\bibfnamefont {J.}~\bibnamefont
  {Luo}} \emph {et~al.} (\bibinfo {collaboration} {TianQin}),\ }\href {\doibase
  10.1088/0264-9381/33/3/035010} {\bibfield  {journal} {\bibinfo  {journal}
  {Class. Quant. Grav.}\ }\textbf {\bibinfo {volume} {33}},\ \bibinfo {pages}
  {035010} (\bibinfo {year} {2016})},\ \Eprint
  {http://arxiv.org/abs/1512.02076} {arXiv:1512.02076 [astro-ph.IM]}
  \BibitemShut {NoStop}%
\bibitem [{\citenamefont {Corbin}\ and\ \citenamefont
  {Cornish}(2006)}]{Corbin:2005ny}%
  \BibitemOpen
  \bibfield  {author} {\bibinfo {author} {\bibfnamefont {V.}~\bibnamefont
  {Corbin}}\ and\ \bibinfo {author} {\bibfnamefont {N.~J.}\ \bibnamefont
  {Cornish}},\ }\href {\doibase 10.1088/0264-9381/23/7/014} {\bibfield
  {journal} {\bibinfo  {journal} {Class. Quant. Grav.}\ }\textbf {\bibinfo
  {volume} {23}},\ \bibinfo {pages} {2435} (\bibinfo {year} {2006})},\ \Eprint
  {http://arxiv.org/abs/gr-qc/0512039} {arXiv:gr-qc/0512039} \BibitemShut
  {NoStop}%
\bibitem [{\citenamefont {Seto}\ \emph {et~al.}(2001)\citenamefont {Seto},
  \citenamefont {Kawamura},\ and\ \citenamefont {Nakamura}}]{Seto:2001qf}%
  \BibitemOpen
  \bibfield  {author} {\bibinfo {author} {\bibfnamefont {N.}~\bibnamefont
  {Seto}}, \bibinfo {author} {\bibfnamefont {S.}~\bibnamefont {Kawamura}}, \
  and\ \bibinfo {author} {\bibfnamefont {T.}~\bibnamefont {Nakamura}},\ }\href
  {\doibase 10.1103/PhysRevLett.87.221103} {\bibfield  {journal} {\bibinfo
  {journal} {Phys. Rev. Lett.}\ }\textbf {\bibinfo {volume} {87}},\ \bibinfo
  {pages} {221103} (\bibinfo {year} {2001})},\ \Eprint
  {http://arxiv.org/abs/astro-ph/0108011} {arXiv:astro-ph/0108011} \BibitemShut
  {NoStop}%
\bibitem [{\citenamefont {Punturo}\ \emph {et~al.}(2010)\citenamefont {Punturo}
  \emph {et~al.}}]{Punturo:2010zz}%
  \BibitemOpen
  \bibfield  {author} {\bibinfo {author} {\bibfnamefont {M.}~\bibnamefont
  {Punturo}} \emph {et~al.},\ }\href {\doibase 10.1088/0264-9381/27/19/194002}
  {\bibfield  {journal} {\bibinfo  {journal} {Class. Quant. Grav.}\ }\textbf
  {\bibinfo {volume} {27}},\ \bibinfo {pages} {194002} (\bibinfo {year}
  {2010})}\BibitemShut {NoStop}%
\bibitem [{\citenamefont {Abbott}\ \emph {et~al.}(2017)\citenamefont {Abbott}
  \emph {et~al.}}]{LIGOScientific:2016wof}%
  \BibitemOpen
  \bibfield  {author} {\bibinfo {author} {\bibfnamefont {B.~P.}\ \bibnamefont
  {Abbott}} \emph {et~al.} (\bibinfo {collaboration} {LIGO Scientific}),\
  }\href {\doibase 10.1088/1361-6382/aa51f4} {\bibfield  {journal} {\bibinfo
  {journal} {Class. Quant. Grav.}\ }\textbf {\bibinfo {volume} {34}},\ \bibinfo
  {pages} {044001} (\bibinfo {year} {2017})},\ \Eprint
  {http://arxiv.org/abs/1607.08697} {arXiv:1607.08697 [astro-ph.IM]}
  \BibitemShut {NoStop}%
\bibitem [{\citenamefont {El-Neaj}\ \emph {et~al.}(2020)\citenamefont {El-Neaj}
  \emph {et~al.}}]{AEDGE:2019nxb}%
  \BibitemOpen
  \bibfield  {author} {\bibinfo {author} {\bibfnamefont {Y.~A.}\ \bibnamefont
  {El-Neaj}} \emph {et~al.} (\bibinfo {collaboration} {AEDGE}),\ }\href
  {\doibase 10.1140/epjqt/s40507-020-0080-0} {\bibfield  {journal} {\bibinfo
  {journal} {EPJ Quant. Technol.}\ }\textbf {\bibinfo {volume} {7}},\ \bibinfo
  {pages} {6} (\bibinfo {year} {2020})},\ \Eprint
  {http://arxiv.org/abs/1908.00802} {arXiv:1908.00802 [gr-qc]} \BibitemShut
  {NoStop}%
\bibitem [{\citenamefont {Ahmed}\ \emph
  {et~al.}(2024{\natexlab{a}})\citenamefont {Ahmed}, \citenamefont {Chowdhury},
  \citenamefont {Nasri},\ and\ \citenamefont {Saad}}]{Ahmed:2023pjl}%
  \BibitemOpen
  \bibfield  {author} {\bibinfo {author} {\bibfnamefont {W.}~\bibnamefont
  {Ahmed}}, \bibinfo {author} {\bibfnamefont {T.~A.}\ \bibnamefont
  {Chowdhury}}, \bibinfo {author} {\bibfnamefont {S.}~\bibnamefont {Nasri}}, \
  and\ \bibinfo {author} {\bibfnamefont {S.}~\bibnamefont {Saad}},\ }\href
  {\doibase 10.1103/PhysRevD.109.015008} {\bibfield  {journal} {\bibinfo
  {journal} {Phys. Rev. D}\ }\textbf {\bibinfo {volume} {109}},\ \bibinfo
  {pages} {015008} (\bibinfo {year} {2024}{\natexlab{a}})},\ \Eprint
  {http://arxiv.org/abs/2308.13248} {arXiv:2308.13248 [hep-ph]} \BibitemShut
  {NoStop}%
\bibitem [{\citenamefont {Ahmed}\ \emph
  {et~al.}(2024{\natexlab{b}})\citenamefont {Ahmed}, \citenamefont {Rehman},\
  and\ \citenamefont {Zubair}}]{Ahmed:2023rky}%
  \BibitemOpen
  \bibfield  {author} {\bibinfo {author} {\bibfnamefont {W.}~\bibnamefont
  {Ahmed}}, \bibinfo {author} {\bibfnamefont {M.~U.}\ \bibnamefont {Rehman}}, \
  and\ \bibinfo {author} {\bibfnamefont {U.}~\bibnamefont {Zubair}},\ }\href
  {\doibase 10.1088/1475-7516/2024/01/049} {\bibfield  {journal} {\bibinfo
  {journal} {JCAP}\ }\textbf {\bibinfo {volume} {01}},\ \bibinfo {pages} {049}
  (\bibinfo {year} {2024}{\natexlab{b}})},\ \Eprint
  {http://arxiv.org/abs/2308.09125} {arXiv:2308.09125 [hep-ph]} \BibitemShut
  {NoStop}%
\bibitem [{\citenamefont {King}\ \emph {et~al.}(2024)\citenamefont {King},
  \citenamefont {Leontaris},\ and\ \citenamefont {Zhou}}]{King:2023wkm}%
  \BibitemOpen
  \bibfield  {author} {\bibinfo {author} {\bibfnamefont {S.~F.}\ \bibnamefont
  {King}}, \bibinfo {author} {\bibfnamefont {G.~K.}\ \bibnamefont {Leontaris}},
  \ and\ \bibinfo {author} {\bibfnamefont {Y.-L.}\ \bibnamefont {Zhou}},\
  }\href {\doibase 10.1007/JHEP03(2024)006} {\bibfield  {journal} {\bibinfo
  {journal} {JHEP}\ }\textbf {\bibinfo {volume} {03}},\ \bibinfo {pages} {006}
  (\bibinfo {year} {2024})},\ \Eprint {http://arxiv.org/abs/2311.11857}
  {arXiv:2311.11857 [hep-ph]} \BibitemShut {NoStop}%
\bibitem [{\citenamefont {Lazarides}\ \emph {et~al.}(2024)\citenamefont
  {Lazarides}, \citenamefont {Maji}, \citenamefont {Moursy},\ and\
  \citenamefont {Shafi}}]{Lazarides:2023rqf}%
  \BibitemOpen
  \bibfield  {author} {\bibinfo {author} {\bibfnamefont {G.}~\bibnamefont
  {Lazarides}}, \bibinfo {author} {\bibfnamefont {R.}~\bibnamefont {Maji}},
  \bibinfo {author} {\bibfnamefont {A.}~\bibnamefont {Moursy}}, \ and\ \bibinfo
  {author} {\bibfnamefont {Q.}~\bibnamefont {Shafi}},\ }\href {\doibase
  10.1088/1475-7516/2024/03/006} {\bibfield  {journal} {\bibinfo  {journal}
  {JCAP}\ }\textbf {\bibinfo {volume} {03}},\ \bibinfo {pages} {006} (\bibinfo
  {year} {2024})},\ \Eprint {http://arxiv.org/abs/2308.07094} {arXiv:2308.07094
  [hep-ph]} \BibitemShut {NoStop}%
\bibitem [{\citenamefont {Lazarides}\ \emph
  {et~al.}(2023{\natexlab{b}})\citenamefont {Lazarides}, \citenamefont {Maji},\
  and\ \citenamefont {Shafi}}]{Lazarides:2023ksx}%
  \BibitemOpen
  \bibfield  {author} {\bibinfo {author} {\bibfnamefont {G.}~\bibnamefont
  {Lazarides}}, \bibinfo {author} {\bibfnamefont {R.}~\bibnamefont {Maji}}, \
  and\ \bibinfo {author} {\bibfnamefont {Q.}~\bibnamefont {Shafi}},\ }\href
  {\doibase 10.1103/PhysRevD.108.095041} {\bibfield  {journal} {\bibinfo
  {journal} {Phys. Rev. D}\ }\textbf {\bibinfo {volume} {108}},\ \bibinfo
  {pages} {095041} (\bibinfo {year} {2023}{\natexlab{b}})},\ \Eprint
  {http://arxiv.org/abs/2306.17788} {arXiv:2306.17788 [hep-ph]} \BibitemShut
  {NoStop}%
\bibitem [{\citenamefont {Buchmuller}\ \emph {et~al.}(2023)\citenamefont
  {Buchmuller}, \citenamefont {Domcke},\ and\ \citenamefont
  {Schmitz}}]{Buchmuller:2023aus}%
  \BibitemOpen
  \bibfield  {author} {\bibinfo {author} {\bibfnamefont {W.}~\bibnamefont
  {Buchmuller}}, \bibinfo {author} {\bibfnamefont {V.}~\bibnamefont {Domcke}},
  \ and\ \bibinfo {author} {\bibfnamefont {K.}~\bibnamefont {Schmitz}},\ }\href
  {\doibase 10.1088/1475-7516/2023/11/020} {\bibfield  {journal} {\bibinfo
  {journal} {JCAP}\ }\textbf {\bibinfo {volume} {11}},\ \bibinfo {pages} {020}
  (\bibinfo {year} {2023})},\ \Eprint {http://arxiv.org/abs/2307.04691}
  {arXiv:2307.04691 [hep-ph]} \BibitemShut {NoStop}%
\bibitem [{\citenamefont {Antusch}\ \emph {et~al.}(2023)\citenamefont
  {Antusch}, \citenamefont {Hinze}, \citenamefont {Saad},\ and\ \citenamefont
  {Steiner}}]{Antusch:2023zjk}%
  \BibitemOpen
  \bibfield  {author} {\bibinfo {author} {\bibfnamefont {S.}~\bibnamefont
  {Antusch}}, \bibinfo {author} {\bibfnamefont {K.}~\bibnamefont {Hinze}},
  \bibinfo {author} {\bibfnamefont {S.}~\bibnamefont {Saad}}, \ and\ \bibinfo
  {author} {\bibfnamefont {J.}~\bibnamefont {Steiner}},\ }\href {\doibase
  10.1103/PhysRevD.108.095053} {\bibfield  {journal} {\bibinfo  {journal}
  {Phys. Rev. D}\ }\textbf {\bibinfo {volume} {108}},\ \bibinfo {pages}
  {095053} (\bibinfo {year} {2023})},\ \Eprint
  {http://arxiv.org/abs/2307.04595} {arXiv:2307.04595 [hep-ph]} \BibitemShut
  {NoStop}%
\bibitem [{\citenamefont {Fu}\ \emph {et~al.}(2024)\citenamefont {Fu},
  \citenamefont {King}, \citenamefont {Marsili}, \citenamefont {Pascoli},
  \citenamefont {Turner},\ and\ \citenamefont {Zhou}}]{Fu:2023mdu}%
  \BibitemOpen
  \bibfield  {author} {\bibinfo {author} {\bibfnamefont {B.}~\bibnamefont
  {Fu}}, \bibinfo {author} {\bibfnamefont {S.~F.}\ \bibnamefont {King}},
  \bibinfo {author} {\bibfnamefont {L.}~\bibnamefont {Marsili}}, \bibinfo
  {author} {\bibfnamefont {S.}~\bibnamefont {Pascoli}}, \bibinfo {author}
  {\bibfnamefont {J.}~\bibnamefont {Turner}}, \ and\ \bibinfo {author}
  {\bibfnamefont {Y.-L.}\ \bibnamefont {Zhou}},\ }\href {\doibase
  10.1103/PhysRevD.109.055025} {\bibfield  {journal} {\bibinfo  {journal}
  {Phys. Rev. D}\ }\textbf {\bibinfo {volume} {109}},\ \bibinfo {pages}
  {055025} (\bibinfo {year} {2024})},\ \Eprint
  {http://arxiv.org/abs/2308.05799} {arXiv:2308.05799 [hep-ph]} \BibitemShut
  {NoStop}%
\bibitem [{\citenamefont {Vagnozzi}(2023)}]{Vagnozzi:2023lwo}%
  \BibitemOpen
  \bibfield  {author} {\bibinfo {author} {\bibfnamefont {S.}~\bibnamefont
  {Vagnozzi}},\ }\href {\doibase 10.1016/j.jheap.2023.07.001} {\bibfield
  {journal} {\bibinfo  {journal} {JHEAp}\ }\textbf {\bibinfo {volume} {39}},\
  \bibinfo {pages} {81} (\bibinfo {year} {2023})},\ \Eprint
  {http://arxiv.org/abs/2306.16912} {arXiv:2306.16912 [astro-ph.CO]}
  \BibitemShut {NoStop}%
\bibitem [{\citenamefont {Vagnozzi}(2021)}]{Vagnozzi:2020gtf}%
  \BibitemOpen
  \bibfield  {author} {\bibinfo {author} {\bibfnamefont {S.}~\bibnamefont
  {Vagnozzi}},\ }\href {\doibase 10.1093/mnrasl/slaa203} {\bibfield  {journal}
  {\bibinfo  {journal} {Mon. Not. Roy. Astron. Soc.}\ }\textbf {\bibinfo
  {volume} {502}},\ \bibinfo {pages} {L11} (\bibinfo {year} {2021})},\ \Eprint
  {http://arxiv.org/abs/2009.13432} {arXiv:2009.13432 [astro-ph.CO]}
  \BibitemShut {NoStop}%
\bibitem [{\citenamefont {Buchmuller}(2021)}]{Buchmuller:2021dtt}%
  \BibitemOpen
  \bibfield  {author} {\bibinfo {author} {\bibfnamefont {W.}~\bibnamefont
  {Buchmuller}},\ }\href {\doibase 10.1007/JHEP04(2021)168} {\bibfield
  {journal} {\bibinfo  {journal} {JHEP}\ }\textbf {\bibinfo {volume} {04}},\
  \bibinfo {pages} {168} (\bibinfo {year} {2021})},\ \Eprint
  {http://arxiv.org/abs/2102.08923} {arXiv:2102.08923 [hep-ph]} \BibitemShut
  {NoStop}%
\bibitem [{\citenamefont {Buchmuller}\ \emph {et~al.}(2021)\citenamefont
  {Buchmuller}, \citenamefont {Domcke},\ and\ \citenamefont
  {Schmitz}}]{Buchmuller:2021mbb}%
  \BibitemOpen
  \bibfield  {author} {\bibinfo {author} {\bibfnamefont {W.}~\bibnamefont
  {Buchmuller}}, \bibinfo {author} {\bibfnamefont {V.}~\bibnamefont {Domcke}},
  \ and\ \bibinfo {author} {\bibfnamefont {K.}~\bibnamefont {Schmitz}},\ }\href
  {\doibase 10.1088/1475-7516/2021/12/006} {\bibfield  {journal} {\bibinfo
  {journal} {JCAP}\ }\textbf {\bibinfo {volume} {12}},\ \bibinfo {pages} {006}
  (\bibinfo {year} {2021})},\ \Eprint {http://arxiv.org/abs/2107.04578}
  {arXiv:2107.04578 [hep-ph]} \BibitemShut {NoStop}%
\bibitem [{\citenamefont {Masoud}\ \emph {et~al.}(2021)\citenamefont {Masoud},
  \citenamefont {Rehman},\ and\ \citenamefont {Shafi}}]{Masoud:2021prr}%
  \BibitemOpen
  \bibfield  {author} {\bibinfo {author} {\bibfnamefont {M.~A.}\ \bibnamefont
  {Masoud}}, \bibinfo {author} {\bibfnamefont {M.~U.}\ \bibnamefont {Rehman}},
  \ and\ \bibinfo {author} {\bibfnamefont {Q.}~\bibnamefont {Shafi}},\ }\href
  {\doibase 10.1088/1475-7516/2021/11/022} {\bibfield  {journal} {\bibinfo
  {journal} {JCAP}\ }\textbf {\bibinfo {volume} {11}},\ \bibinfo {pages} {022}
  (\bibinfo {year} {2021})},\ \Eprint {http://arxiv.org/abs/2107.09689}
  {arXiv:2107.09689 [hep-ph]} \BibitemShut {NoStop}%
\bibitem [{\citenamefont {Ahmed}\ \emph {et~al.}(2022)\citenamefont {Ahmed},
  \citenamefont {Junaid}, \citenamefont {Nasri},\ and\ \citenamefont
  {Zubair}}]{Ahmed:2022rwy}%
  \BibitemOpen
  \bibfield  {author} {\bibinfo {author} {\bibfnamefont {W.}~\bibnamefont
  {Ahmed}}, \bibinfo {author} {\bibfnamefont {M.}~\bibnamefont {Junaid}},
  \bibinfo {author} {\bibfnamefont {S.}~\bibnamefont {Nasri}}, \ and\ \bibinfo
  {author} {\bibfnamefont {U.}~\bibnamefont {Zubair}},\ }\href {\doibase
  10.1103/PhysRevD.105.115008} {\bibfield  {journal} {\bibinfo  {journal}
  {Phys. Rev. D}\ }\textbf {\bibinfo {volume} {105}},\ \bibinfo {pages}
  {115008} (\bibinfo {year} {2022})},\ \Eprint
  {http://arxiv.org/abs/2202.06216} {arXiv:2202.06216 [hep-ph]} \BibitemShut
  {NoStop}%
\bibitem [{\citenamefont {Pallis}(2024)}]{Pallis:2024mip}%
  \BibitemOpen
  \bibfield  {author} {\bibinfo {author} {\bibfnamefont {C.}~\bibnamefont
  {Pallis}},\ }\href@noop {} {\  (\bibinfo {year} {2024})},\ \Eprint
  {http://arxiv.org/abs/2403.09385} {arXiv:2403.09385 [hep-ph]} \BibitemShut
  {NoStop}%
\bibitem [{\citenamefont {Tanabashi}\ \emph {et~al.}(2018)\citenamefont
  {Tanabashi} \emph {et~al.}}]{ParticleDataGroup:2018ovx}%
  \BibitemOpen
  \bibfield  {author} {\bibinfo {author} {\bibfnamefont {M.}~\bibnamefont
  {Tanabashi}} \emph {et~al.} (\bibinfo {collaboration} {Particle Data
  Group}),\ }\href {\doibase 10.1103/PhysRevD.98.030001} {\bibfield  {journal}
  {\bibinfo  {journal} {Phys. Rev. D}\ }\textbf {\bibinfo {volume} {98}},\
  \bibinfo {pages} {030001} (\bibinfo {year} {2018})}\BibitemShut {NoStop}%
\bibitem [{\citenamefont {Antoniadis}\ \emph
  {et~al.}(2023{\natexlab{b}})\citenamefont {Antoniadis} \emph
  {et~al.}}]{EPTA:2023xxk}%
  \BibitemOpen
  \bibfield  {author} {\bibinfo {author} {\bibfnamefont {J.}~\bibnamefont
  {Antoniadis}} \emph {et~al.} (\bibinfo {collaboration} {EPTA}),\ }\href@noop
  {} {\  (\bibinfo {year} {2023}{\natexlab{b}})},\ \Eprint
  {http://arxiv.org/abs/2306.16227} {arXiv:2306.16227 [astro-ph.CO]}
  \BibitemShut {NoStop}%
\bibitem [{\citenamefont {Hill}\ \emph {et~al.}(1988)\citenamefont {Hill},
  \citenamefont {Hodges},\ and\ \citenamefont {Turner}}]{Hill:1987qx}%
  \BibitemOpen
  \bibfield  {author} {\bibinfo {author} {\bibfnamefont {C.~T.}\ \bibnamefont
  {Hill}}, \bibinfo {author} {\bibfnamefont {H.~M.}\ \bibnamefont {Hodges}}, \
  and\ \bibinfo {author} {\bibfnamefont {M.~S.}\ \bibnamefont {Turner}},\
  }\href {\doibase 10.1103/PhysRevD.37.263} {\bibfield  {journal} {\bibinfo
  {journal} {Phys. Rev. D}\ }\textbf {\bibinfo {volume} {37}},\ \bibinfo
  {pages} {263} (\bibinfo {year} {1988})}\BibitemShut {NoStop}%
\bibitem [{\citenamefont {Pati}\ and\ \citenamefont
  {Salam}(1973)}]{Pati:1973rp}%
  \BibitemOpen
  \bibfield  {author} {\bibinfo {author} {\bibfnamefont {J.~C.}\ \bibnamefont
  {Pati}}\ and\ \bibinfo {author} {\bibfnamefont {A.}~\bibnamefont {Salam}},\
  }\href {\doibase 10.1103/PhysRevLett.31.661} {\bibfield  {journal} {\bibinfo
  {journal} {Phys. Rev. Lett.}\ }\textbf {\bibinfo {volume} {31}},\ \bibinfo
  {pages} {661} (\bibinfo {year} {1973})}\BibitemShut {NoStop}%
\bibitem [{\citenamefont {Abbott}\ \emph {et~al.}(2021)\citenamefont {Abbott}
  \emph {et~al.}}]{LIGOScientific:2021nrg}%
  \BibitemOpen
  \bibfield  {author} {\bibinfo {author} {\bibfnamefont {R.}~\bibnamefont
  {Abbott}} \emph {et~al.} (\bibinfo {collaboration} {LIGO Scientific, Virgo,
  KAGRA}),\ }\href {\doibase 10.1103/PhysRevLett.126.241102} {\bibfield
  {journal} {\bibinfo  {journal} {Phys. Rev. Lett.}\ }\textbf {\bibinfo
  {volume} {126}},\ \bibinfo {pages} {241102} (\bibinfo {year} {2021})},\
  \Eprint {http://arxiv.org/abs/2101.12248} {arXiv:2101.12248 [gr-qc]}
  \BibitemShut {NoStop}%
\bibitem [{\citenamefont {Abbott}\ and\ \citenamefont
  {Wise}(1980)}]{Abbott:1980zj}%
  \BibitemOpen
  \bibfield  {author} {\bibinfo {author} {\bibfnamefont {L.~F.}\ \bibnamefont
  {Abbott}}\ and\ \bibinfo {author} {\bibfnamefont {M.~B.}\ \bibnamefont
  {Wise}},\ }\href {\doibase 10.1103/PhysRevD.22.2208} {\bibfield  {journal}
  {\bibinfo  {journal} {Phys. Rev. D}\ }\textbf {\bibinfo {volume} {22}},\
  \bibinfo {pages} {2208} (\bibinfo {year} {1980})}\BibitemShut {NoStop}%
\bibitem [{\citenamefont {Munoz}(1986)}]{Munoz:1986kq}%
  \BibitemOpen
  \bibfield  {author} {\bibinfo {author} {\bibfnamefont {C.}~\bibnamefont
  {Munoz}},\ }\href {\doibase 10.1016/0370-2693(86)90013-4} {\bibfield
  {journal} {\bibinfo  {journal} {Phys. Lett. B}\ }\textbf {\bibinfo {volume}
  {177}},\ \bibinfo {pages} {55} (\bibinfo {year} {1986})}\BibitemShut
  {NoStop}%
\bibitem [{\citenamefont {Nihei}\ and\ \citenamefont
  {Arafune}(1995)}]{Nihei:1994tx}%
  \BibitemOpen
  \bibfield  {author} {\bibinfo {author} {\bibfnamefont {T.}~\bibnamefont
  {Nihei}}\ and\ \bibinfo {author} {\bibfnamefont {J.}~\bibnamefont
  {Arafune}},\ }\href {\doibase 10.1143/PTP.93.665} {\bibfield  {journal}
  {\bibinfo  {journal} {Prog. Theor. Phys.}\ }\textbf {\bibinfo {volume}
  {93}},\ \bibinfo {pages} {665} (\bibinfo {year} {1995})},\ \Eprint
  {http://arxiv.org/abs/hep-ph/9412325} {arXiv:hep-ph/9412325} \BibitemShut
  {NoStop}%
\bibitem [{\citenamefont {Aoki}\ \emph {et~al.}(2017)\citenamefont {Aoki},
  \citenamefont {Izubuchi}, \citenamefont {Shintani},\ and\ \citenamefont
  {Soni}}]{Aoki:2017puj}%
  \BibitemOpen
  \bibfield  {author} {\bibinfo {author} {\bibfnamefont {Y.}~\bibnamefont
  {Aoki}}, \bibinfo {author} {\bibfnamefont {T.}~\bibnamefont {Izubuchi}},
  \bibinfo {author} {\bibfnamefont {E.}~\bibnamefont {Shintani}}, \ and\
  \bibinfo {author} {\bibfnamefont {A.}~\bibnamefont {Soni}},\ }\href {\doibase
  10.1103/PhysRevD.96.014506} {\bibfield  {journal} {\bibinfo  {journal} {Phys.
  Rev. D}\ }\textbf {\bibinfo {volume} {96}},\ \bibinfo {pages} {014506}
  (\bibinfo {year} {2017})},\ \Eprint {http://arxiv.org/abs/1705.01338}
  {arXiv:1705.01338 [hep-lat]} \BibitemShut {NoStop}%
\bibitem [{\citenamefont {Abe}\ \emph {et~al.}(2017)\citenamefont {Abe} \emph
  {et~al.}}]{Super-Kamiokande:2016exg}%
  \BibitemOpen
  \bibfield  {author} {\bibinfo {author} {\bibfnamefont {K.}~\bibnamefont
  {Abe}} \emph {et~al.} (\bibinfo {collaboration} {Super-Kamiokande}),\ }\href
  {\doibase 10.1103/PhysRevD.95.012004} {\bibfield  {journal} {\bibinfo
  {journal} {Phys. Rev. D}\ }\textbf {\bibinfo {volume} {95}},\ \bibinfo
  {pages} {012004} (\bibinfo {year} {2017})},\ \Eprint
  {http://arxiv.org/abs/1610.03597} {arXiv:1610.03597 [hep-ex]} \BibitemShut
  {NoStop}%
\bibitem [{\citenamefont {Abe}\ \emph {et~al.}(2014{\natexlab{a}})\citenamefont
  {Abe} \emph {et~al.}}]{Super-Kamiokande:2013rwg}%
  \BibitemOpen
  \bibfield  {author} {\bibinfo {author} {\bibfnamefont {K.}~\bibnamefont
  {Abe}} \emph {et~al.} (\bibinfo {collaboration} {Super-Kamiokande}),\ }\href
  {\doibase 10.1103/PhysRevLett.113.121802} {\bibfield  {journal} {\bibinfo
  {journal} {Phys. Rev. Lett.}\ }\textbf {\bibinfo {volume} {113}},\ \bibinfo
  {pages} {121802} (\bibinfo {year} {2014}{\natexlab{a}})},\ \Eprint
  {http://arxiv.org/abs/1305.4391} {arXiv:1305.4391 [hep-ex]} \BibitemShut
  {NoStop}%
\bibitem [{\citenamefont {Takhistov}(2016)}]{Takhistov:2016eqm}%
  \BibitemOpen
  \bibfield  {author} {\bibinfo {author} {\bibfnamefont {V.}~\bibnamefont
  {Takhistov}} (\bibinfo {collaboration} {Super-Kamiokande}),\ }in\ \href@noop
  {} {\emph {\bibinfo {booktitle} {{51st Rencontres de Moriond on EW
  Interactions and Unified Theories}}}}\ (\bibinfo {year} {2016})\ pp.\
  \bibinfo {pages} {437--444},\ \Eprint {http://arxiv.org/abs/1605.03235}
  {arXiv:1605.03235 [hep-ex]} \BibitemShut {NoStop}%
\bibitem [{\citenamefont {Regis}\ \emph {et~al.}(2012)\citenamefont {Regis}
  \emph {et~al.}}]{Super-Kamiokande:2012zik}%
  \BibitemOpen
  \bibfield  {author} {\bibinfo {author} {\bibfnamefont {C.}~\bibnamefont
  {Regis}} \emph {et~al.} (\bibinfo {collaboration} {Super-Kamiokande}),\
  }\href {\doibase 10.1103/PhysRevD.86.012006} {\bibfield  {journal} {\bibinfo
  {journal} {Phys. Rev. D}\ }\textbf {\bibinfo {volume} {86}},\ \bibinfo
  {pages} {012006} (\bibinfo {year} {2012})},\ \Eprint
  {http://arxiv.org/abs/1205.6538} {arXiv:1205.6538 [hep-ex]} \BibitemShut
  {NoStop}%
\bibitem [{\citenamefont {Abe}\ \emph {et~al.}(2014{\natexlab{b}})\citenamefont
  {Abe} \emph {et~al.}}]{Super-Kamiokande:2014otb}%
  \BibitemOpen
  \bibfield  {author} {\bibinfo {author} {\bibfnamefont {K.}~\bibnamefont
  {Abe}} \emph {et~al.} (\bibinfo {collaboration} {Super-Kamiokande}),\ }\href
  {\doibase 10.1103/PhysRevD.90.072005} {\bibfield  {journal} {\bibinfo
  {journal} {Phys. Rev. D}\ }\textbf {\bibinfo {volume} {90}},\ \bibinfo
  {pages} {072005} (\bibinfo {year} {2014}{\natexlab{b}})},\ \Eprint
  {http://arxiv.org/abs/1408.1195} {arXiv:1408.1195 [hep-ex]} \BibitemShut
  {NoStop}%
\bibitem [{\citenamefont {Kobayashi}\ \emph {et~al.}(2005)\citenamefont
  {Kobayashi} \emph {et~al.}}]{Super-Kamiokande:2005lev}%
  \BibitemOpen
  \bibfield  {author} {\bibinfo {author} {\bibfnamefont {K.}~\bibnamefont
  {Kobayashi}} \emph {et~al.} (\bibinfo {collaboration} {Super-Kamiokande}),\
  }\href {\doibase 10.1103/PhysRevD.72.052007} {\bibfield  {journal} {\bibinfo
  {journal} {Phys. Rev. D}\ }\textbf {\bibinfo {volume} {72}},\ \bibinfo
  {pages} {052007} (\bibinfo {year} {2005})},\ \Eprint
  {http://arxiv.org/abs/hep-ex/0502026} {arXiv:hep-ex/0502026} \BibitemShut
  {NoStop}%
\bibitem [{\citenamefont {Aghanim}\ \emph {et~al.}(2020)\citenamefont {Aghanim}
  \emph {et~al.}}]{Planck:2018vyg}%
  \BibitemOpen
  \bibfield  {author} {\bibinfo {author} {\bibfnamefont {N.}~\bibnamefont
  {Aghanim}} \emph {et~al.} (\bibinfo {collaboration} {Planck}),\ }\href
  {\doibase 10.1051/0004-6361/201833910} {\bibfield  {journal} {\bibinfo
  {journal} {Astron. Astrophys.}\ }\textbf {\bibinfo {volume} {641}},\ \bibinfo
  {pages} {A6} (\bibinfo {year} {2020})},\ \bibinfo {note} {[Erratum:
  Astron.Astrophys. 652, C4 (2021)]},\ \Eprint
  {http://arxiv.org/abs/1807.06209} {arXiv:1807.06209 [astro-ph.CO]}
  \BibitemShut {NoStop}%
\bibitem [{\citenamefont {Smits}\ \emph {et~al.}(2009)\citenamefont {Smits},
  \citenamefont {Kramer}, \citenamefont {Stappers}, \citenamefont {Lorimer},
  \citenamefont {Cordes},\ and\ \citenamefont {Faulkner}}]{Smits:2008cf}%
  \BibitemOpen
  \bibfield  {author} {\bibinfo {author} {\bibfnamefont {R.}~\bibnamefont
  {Smits}}, \bibinfo {author} {\bibfnamefont {M.}~\bibnamefont {Kramer}},
  \bibinfo {author} {\bibfnamefont {B.}~\bibnamefont {Stappers}}, \bibinfo
  {author} {\bibfnamefont {D.~R.}\ \bibnamefont {Lorimer}}, \bibinfo {author}
  {\bibfnamefont {J.}~\bibnamefont {Cordes}}, \ and\ \bibinfo {author}
  {\bibfnamefont {A.}~\bibnamefont {Faulkner}},\ }\href {\doibase
  10.1051/0004-6361:200810383} {\bibfield  {journal} {\bibinfo  {journal}
  {Astron. Astrophys.}\ }\textbf {\bibinfo {volume} {493}},\ \bibinfo {pages}
  {1161} (\bibinfo {year} {2009})},\ \Eprint {http://arxiv.org/abs/0811.0211}
  {arXiv:0811.0211 [astro-ph]} \BibitemShut {NoStop}%
\end{thebibliography}%
